\documentclass{aa}  
\usepackage[nolist]{acronym}

\usepackage{graphicx}
\usepackage{txfonts}
\usepackage{natbib}
\usepackage{booktabs}
\usepackage{color}
\usepackage{colortbl}
\usepackage{hyperref}
\usepackage{subcaption}
\usepackage{nccmath}
\usepackage[flushleft]{threeparttable}
\hypersetup{colorlinks=true,allcolors=[rgb]{0,0,0.8}}
\definecolor{grijs}{gray}{0.40}
\definecolor{lichtgrijs}{gray}{0.90}
\makeatletter
\renewcommand*\aa@pageof{, page \thepage{} of \pageref*{LastPage}}

\newacro{etwo}[EPIC 2202]{EPIC~220208795}

\newcommand{\rsun}{$R_{\odot}$}
\newcommand{\teff}{$T_{\rm eff}$}
\newcommand{\logg}{\ensuremath{\log g}}
\newcommand{\Msun}{$M_{\odot}$}
\newcommand{\logl}{log($L/L_{\odot}$)}

\makeatother
\begin{document} 

\title{K2 Discovery of a Circumsecondary Disk \\Transiting EPIC 220208795}

   \author{L. van der Kamp
          \inst{1}
          \and
          D.M. van Dam\inst{1}
          \and
          M.A. Kenworthy\inst{1}
          \and
          E.E. Mamajek\inst{2,3}
          \and
          G. Pojma\'nski\inst{4}
          }
   \institute{Leiden Observatory, University of Leiden,
   PO Box 9513, 2300 RA Leiden, The Netherlands\\
   \email{lvanderkamp@strw.leidenuniv.nl}
    \and
    Jet Propulsion Laboratory, M/S 321-100, 4800 Oak Grove Drive, Pasadena, CA 91109, USA
    \and
    Department of Physics \& Astronomy, University of Rochester, P.O. Box 270171, Rochester, NY 14627, USA
    \and
    Astronomical Observatory, University of Warsaw, Al. Ujazdowskie 4, 00-478 Warszawa, Poland}

   \date{Submitted 2021 Aug 22; accepted 2021 Oct 23}

  \abstract
   {Observations of the star EPIC 220208795 (2MASS J01105556+0018507) reveal a single, deep and asymmetric eclipse, which we hypothesize is due to an eclipsing companion surrounded by a tilted and inclined opaque disk, similar to those seen around V928~Tau and EPIC 204376071.}
   {We aim to derive physical parameters of the disk and orbital parameters for the companion around the primary star.}
   {The modeling is carried out using a modified version of the \texttt{python} package \texttt{pyPplusS}, and optimization is done using \texttt{emcee}.
     The period analysis makes use of photometry from ground-based surveys, where we perform a period folding search for other possible eclipses by the disk.
   Parameters obtained by the best model fits are used to obtain the parameter space of the orbital parameters, while the most likely period obtained is used to constrain these parameters.}
   {The best model has an opaque disk with a radius of $1.14\pm0.03$ \rsun, an impact parameter of $0.61\pm0.02$ \rsun, an inclination of $77.01^{\circ}\pm0.03^{\circ}$, a tilt of $36.81^{\circ}\pm0.05^{\circ}$ and a transverse velocity of $77.45\pm0.05$ km\,s$^{-1}$.
    The two most likely periods are $\sim 290$ days and $\sim 236$ days, corresponding to an eccentricity of $\sim 0.7$, allowing us to make predictions for the epochs of the next eclipses.
   All models with tilted and inclined disks result in a minimum derived eccentricity of 0.3, which in combination with the two other known small transiting disk candidates V928~Tau and EPIC 204376071, suggest that there may be a common origin for their eccentric orbits.}
   {}


   \maketitle
%

\section{Introduction}

Advances in high precision photometry has allowed astronomers to continuously monitor the apparent brightness of stars, and some of these stars exhibit deep and irregular eclipses in their apparent brightness over time.
These patterns can come from intrinsic stellar variability \citep{joy_t_1945,lanza_comparing_2007,olah_multiple_2009,handler_asteroseismology_2013} or external objects transiting the star, ranging from exoplanets, to exocomets \citep{Rappaport18,Zieba19} to material in and above circumstellar disks \citep{ansdell_little_2019,Kennedy20} or infalling material onto white dwarfs \citep{Vanderburg15,Gaensicke19}.
Other light curves challenge easy identification of the associated astrophysical processes - two notable cases are Boyajian's star \citep{Boyajian15} and HD~139139 \citep{Rappaport19b}. 
Dust that is believed to come from material in the inner part of circumstellar disks can produce dips in the brightness of stars that are called ``dippers'' of $\sim$10-50\% \citep{2016ApJ...816...69A,2019MNRAS.483.3579A,alencar_accretion_2010,cody_csi_2014,cody_many-faceted_2018}.
Circumstellar disks are a universal feature of star formation and also dictate structure of the planetary system that can be formed by the material inside these disks \citep{williams_protoplanetary_2011}.
The formation of gas planets is thought to be through the accretion of material from circumstellar disks that subsequently passes through a circumplanetary disk.
A tilted and inclined disk around a planet or substellar companion can cause a dip in stellar brightness inconsistent with a transiting exoplanet eclipse.
The projection of such a disk creates an elliptical occulting region and (for non-zero impact parameters) creates an asymmetric eclipse. 

Analyzing such systems can yield insights into the formation mechanisms of planets, especially if the observed system is young, since planets form and grow within protoplanetary disks.
The properties and dynamics of these disks are connected, not only to stellar evolution and the evolution of a circumstellar disk, but also to the properties and evolution of the planet \citep{armitage_dynamics_2011,kley_planet-disk_2012}.
Analysing a transit can reveal the structure of the circumplanetary disk and provide insight into the dynamics and mechanics of ring and moon formation \citep{teachey_hek_2017}.
Examples of disk systems or dusty occultations analysed in previous studies are ``J1407'' (V1400 Cen, \citealt{kenworthy_modeling_2015}), EPIC 204376071 (EPIC 2043, \citealt{rappaport_deep_2019}) and V928~Tau \citep{van_dam_asymmetric_2020}.
In the case of J1407, a ring system is hypothesised to be around a secondary companion occulting the star, whereas the other two systems are of an inclined circular disk around a secondary companion occulting the star. 
An elliptical occulter has recently been suggested for the light curve seen towards the late type giant star labelled VVV-WIT-08 \citep{Smith21} discovered as part of the VISTA Variables in the Via-Lactea \citep[VVV; ][]{Minniti10,Minniti17} survey, although this is closer in diameter to the J1407 occulter, and produced an eclipse that lasted about 200 days.

The \textit{Kepler} mission was launched to determine the frequency of Earth-sized planets in and near the habitable zone of Sun-like stars \citep{borucki_kepler_2010}.
The mission has measured a large number of high-resolution light curves and has discovered a great deal of exoplanets \citep{Thompson18_KeplerDR25}.
After the failure of \textit{Kepler}'s second reaction wheel in May 2013, the repurposed \textit{K2} mission \citep{howell_k2_2014} was able to survey the ecliptic plane, leading to numerous additional discoveries \citep{Mayo18_K2}.
Several single, deep transit events have been identified in the \textit{K2} data \citep{lacourse_single_2018}, and we examine the curves that have been classified as `Deep', looking specifically for asymmetric eclipses as these hint at an elliptical occulter.
\ac{etwo}, also known by the aliases TIC~336889445, 2MASS~J01105556+0018507, SDSS~J011055.57 +001850.6 and Gaia~DR2 2534801707104852864, shows a deep and asymmetric transit event similar to the ones observed in \citet{rappaport_deep_2019,van_dam_asymmetric_2020} and we model this eclipse with a tilted and inclined disk around a companion occulting the star.
Section~\ref{sec:data} gives a description of the telescopes and the photometry obtained from them, and the preliminary analysis on the \textit{K2} light curves to identify the most likely candidates for a tilted and inclined disk transit and why we proceeded with \ac{etwo}.
Section~\ref{sec:analysis} describes the modeling of the asymmetric dip found in the \ac{etwo} light curve and we give a description of the best fits for two separate models.
In Section~\ref{sec:period} we perform a period folding analysis using ground-based survey photometry to determine the most likely period given the photometry available and identify other possible eclipses.
Section~\ref{sec:orbit} contains the orbital analysis of the best fit models, obtaining orbital parameters like the eccentricity and periastron distance.
In Section~\ref{sec:discuss} we discuss our results, point out limitations of our modeling, and compare them to two other disk systems from \cite{rappaport_deep_2019}
and \cite{van_dam_asymmetric_2020}.
Section~\ref{sec:conc} summarizes our findings and presents suggestions for future research.

\section{Data}\label{sec:data}

The entire code used for the analysis and creating the figures in this paper is available at \url{ https://github.com/lizvdkamp/EPIC2202_disk}.

\subsection{\textit{K2}}

The \textit{Kepler} spacecraft is a 0.95 m Schmidt telescope with a 1.4 m diameter primary and a 110 deg$^2$ field of view. 
The prime focus camera has 42 CCDs that are $2200\times1024$ pixels \citep{borucki_kepler_2010}.
The extended \textit{K2} mission observed EPIC 2202 during campaign 8 (2016-Jan-03 to 2016-Mar-23) for a total of 79 days. 
It collected 3840 observations, with a cadence of 30 minutes, in the $K_p$ filter, with a bandpass of 420-900 nm.
Further information is listed in Table \ref{tab:surveys}.

The light curve was extracted using \texttt{EVEREST 2.0}, \citep{luger_everest_2016,luger_update_2018}, which is an open-source pipeline for removing instrumental noise from \textit{K2} data.
It uses a variant of Pixel Level Decorrelation \citep[PLD;][]{deming_spitzer_2015} to reduce systematic errors caused by the \textit{Kepler} spacecraft's pointing error, such as the 6 hour trend in the raw aperture photometry, which compromises its ability to detect small transits.
The light curve reveals a star with low amplitude flux variations (less than 1\%), which were deemed small enough to avoid stellar variability modeling. 

Although, the main science mission for \textit{Kepler} and \textit{K2} was the detection and characterisation of exoplanet transits, the library of light curves collected include many other types of astrophysical phenomena.
To this end, \citet{lacourse_single_2018} manually inspected 238,399 stellar light curves and identified 48 stars that showed deep or unusual eclipses.
Ten stars were observed on two separate campaigns resulting in a total of 53 light curves.
The light curves were examined for asymmetry as this is a tell-tale sign of an elliptical occulter, which hints at circumsecondary disk transits.
Of these targets we determined that \ac{etwo} was the most likely candidate for a tilted and inclined disk system, similar to the transits seen in the case of EPIC 2043 and V928~Tau.
The eclipse of \ac{etwo} lasts for about 7.2 hours with a maximum depth of about 25\%.

\begin{figure*}[ht]
    \centering
    \includegraphics[width=\linewidth]{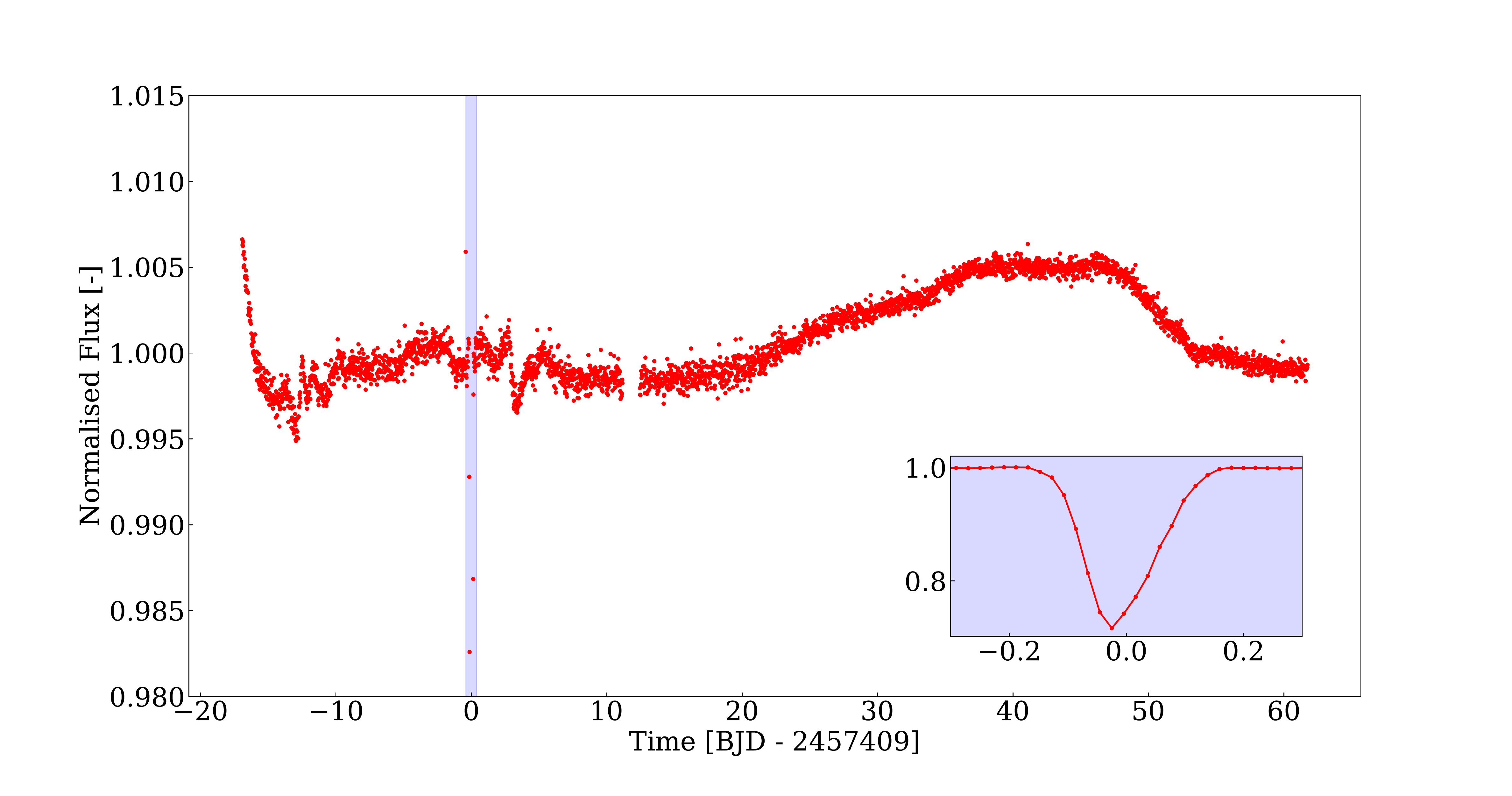}
    \centering
    \caption{\texttt{EVEREST 2.0} light curve of \ac{etwo} with the eclipse centered at BJD 2457409, highlighted in pale blue. Points with non-zero \texttt{QUALITY} values are removed, and the eclipse points have been added back into the light curve (these were incorrectly removed in the pipeline).}
    \label{fig:EVEREST2202}
\end{figure*}

\subsection{TESS} 
The Transiting Exoplanet Survey Satellite (TESS) is a satellite designed to survey for transiting exoplanets among the brightest (and nearest) stars over most of the sky \citep{2015JATIS...1a4003R}.
The TESS satellite orbits the Earth every 13.7 days on a highly elliptical orbit, scanning a sector of the sky spanning $24\times96$ deg$^2$ for a total of two orbits, before moving on to the next sector. 
It captures images at a 2 second (used for guiding), 20 second (for 1 000 bright asteroseismology targets), 120 second (for 200 000 stars that are likely planet hosts) and 30 minute (full frame images) cadences.
The instrument consists of 4 CCDs each with a field of view of $24\times24$ deg$^2$, with a wide band-pass filter from 600-1000 nm (similar to the $I_C$ band) and has a limiting magnitude of about 14-15 mag ($I_C$).
Further information is presented in Table \ref{tab:surveys}.

\begin{figure*}[ht]
    \centering
    \includegraphics[width=\linewidth]{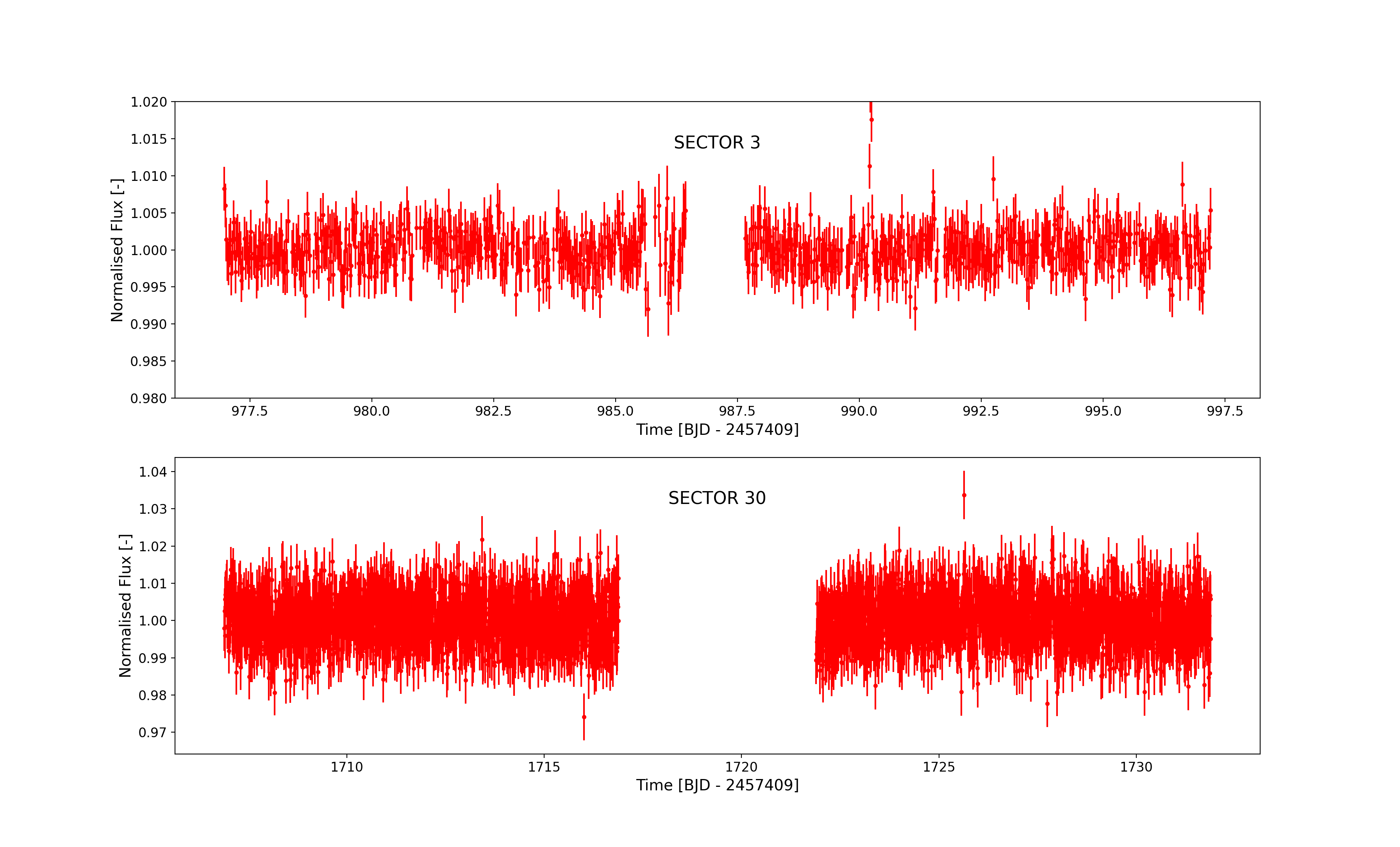}
    \centering
    \caption{The \texttt{eleanor} light curve of EPIC~2202. The upper panel shows Sector 3 and the lower panel shows Sector 30. Note that in Sector 30 there was significantly more scattered light, which explains the larger scatter and the larger absence of data between the two orbits.}
    \label{fig:tess}
\end{figure*}

EPIC 2202 was observed in Sector 3 (2018-Sep-20 to 2018-Oct-18) providing 770 photometric points with a cadence of 30 minutes and in Sector 30 (2020-Sep-22 to 2020-Oct-21) providing 3511 photometric points with a cadence of 10 minutes (see Figure \ref{fig:tess}).
Photometry for TESS has been processed using the \texttt{eleanor} \citep{feinstein_eleanor_2019}, an open-source tool that produces light curves from TESS Full Frame Images.

\subsection{Ground-based Surveys} 

To supplement the space-based data obtained by \textit{K2} and TESS, several archival databases for ground-based surveys were queried for data on EPIC 2202, resulting in three sets of light curves.
This relatively low number is most likely due to the fact that EPIC 2202 is  faint (see Table \ref{tab:Stellarprop}).

The first of these ground-based surveys is the All Sky Automated Survey (ASAS; \citealt{pojmanski_all_1997, asas_2005, asas_2018}).
This is a survey consisting of two observing stations - one in Las
Campanas, Chile and the other on Maui, Hawaii. 
Recently each observatory was equipped with two CCD cameras using $V$ and $I$ filters and commercial f $ = 200$ mm, D $= 100$ mm lenses, although earlier both larger (D $= 250$ mm) and smaller (50-72 mm) lenses were used. Most data are taken with pixel scale of $\approx$ 15". 
ASAS splits the sky into 709 partially overlapping (9x9 deg$^2$ fields, taking on average 150 3-minute  exposures per night, leading to variable cadence of 0.3-2 frames per night.
Depending on the equipment used and the mode of operation, the ASAS limiting magnitude varied between 13.5 and 15.5 mag in V, and the saturation limit was 5.5 to 7.5 mag. 
Precision is around 0.01-0.02 mag for bright stars and below 0.3 mag for the faint ones. 
ASAS photometry is calibrated against the Tycho catalog, and its accuracy is not better than 0.05 mag for bright, non-blended stars.

The second is the All Sky Automated Survey for Supernovae \citep[ASAS-SN; ][]{shappee_man_2014,kochanek_all-sky_2017} which consists of six stations around the globe, with each station hosting four telescopes with a shared mount.
The telescopes consist of a 14-cm aperture telephoto lens with a field of view of approximately $4.5\times4.5$ deg$^2$ and an 8.0'' pixel scale.
Two of the original stations (one in Hawaii and one in Chile) are fitted with $V$ band filters, whereas the additional stations (Chile, Texas, South Africa and China) are fitted with $g$ band filters.
ASAS-SN observes the whole sky every night with a limiting magnitude of about 17 mag in the $V$ and $g$ bands.
The data was obtained through the publicly available Sky Patrol Service\footnote{https://asas-sn.osu.edu}.

The third and final survey is the Asteroid Terrestrial-impact Last Alert System \citep[ATLAS; ][]{heinze_first_2018,tonry_atlas_2018}. 
This is a deep survey with a limiting magnitude of 19 and a precision of around 0.01-0.04 mag.
It consists of two telescopes in Hawaii with a 0.65 m primary mirror and a 0.5 m Schmidt corrector.
Mounted on the telescope is a camera with a 13.5-cm aperture lens, which when combined with the main telescope, provides a field of view of approximately 29 deg$^2$. 
The ATLAS survey uses three special filters $c$ (cyan, 420-650 nm), $o$ (orange, 560-820 nm) and $t$ (tomato, 560-975 nm), that are designed to be differentially sensitive to the silicate colors of stony asteroids.
Data for the ATLAS survey was obtained via their online ``forced photometry'' server\footnote{https://fallingstar-data.com/forcedphot/}.

Further information on the ground-based surveys is presented in Table \ref{tab:surveys} and a light curve of all the photometric time-series data collected is presented in Figure \ref{fig:groundbased}.

\bgroup

\begin{table*}[htb!]
    \centering
    \caption{Overview of the Time-Series Photometric Instruments.}
    \begin{center}
        \begin{tabular}{lcccccccc}
            \toprule
            \toprule
             Survey   & Filter  & $n_{tel}$ & $n_{phot}$ & \begin{tabular}[c]{@{}c@{}}Baseline\\ (days)\end{tabular} & Start Date  & End Date  &  \begin{tabular}[c]{@{}c@{}}Field of View\\ (deg$^2$ cam$^{-1}$)\end{tabular} & \begin{tabular}[c]{@{}c@{}}Pixel-Scale\\ (\arcsec\,pix$^{-1}$)\end{tabular} \\
            \midrule
             ASAS$^a$    & $I$     & 1 - 4     & 85         & 2026    & 27-12-2002  & 14-07-2008     & 4.8, 77.4       & 14.2    \\
                         & $V$     &           & 829        & 7007    & 20-11-2000  & 27-01-2020     &                 &         \\
             ASAS-SN     & $g$     & 12        & 1604       & 1244    & 17-09-2017  & 12-02-2021     & 4.5             & 8.0     \\
                         & $V$     & 8         & 1049       & 2371    & 02-06-2012  & 29-11-2018     &                 &         \\
             ATLAS       & $c$     & 2         & 356        & 2006    & 12-08-2015  & 07-02-2021     & 29.2            & 1.9     \\
                         & $o$     &           & 892        & 1926    & 23-10-2015  & 01-02-2021     &                 &         \\ 
             \textit{K2} & $K_p$   & 1         & 3840       & 79      & 04-01-2016  & 23-03-2016     & 110             & 4.0     \\ 
             TESS        & $I_C^b$ & 1         & 770        & 20      & 24-09-2018  & 14-10-2018     & 576             & 21      \\ 
            \bottomrule
        \end{tabular}
    \end{center}
    \label{tab:surveys}
    \begin{tablenotes}
        \centering
        \item[a] $^a$ In 2000 the survey upgraded to two telescopes in Chile. In 2006 two telescopes in Hawaii were added.
        \item[b] $^b$ Centered on the traditional $I_C$ band, but has a 600-1000 nm bandpass.
      \end{tablenotes}
\end{table*}
\egroup

\subsection{Photometry and Astrometric Data} 

Table \ref{tab:Stellarprop} summarises the available astrometry and photometry for \ac{etwo}. 
The parallax and proper motion are listed from  \textit{Gaia} Early Data Release 3 \citep[\textit{Gaia} EDR3;][]{GaiaEDR3}, and the distance is a geometric estimate based on the {\it Gaia} EDR3 parallax taking into account a prior Galactic model \citep{BailerJones21}.
Mean photometric magnitudes are provided by {\it Gaia} EDR3, 
the Two Micron All Sky Survey \citep[2MASS; ][]{skrutskie_two_2006}, 
and the Sloan Digital Sky Survey Data Release 8 \citep[SDSS DR8;][]{aihara_eighth_2011}. 

\bgroup
\begin{table}[ht]
    \centering
    \captionsetup{justification=centering}
    \caption{ Properties of \ac{etwo}.}
    \begin{tabular}{@{}lcc@{}}
        \toprule
        \toprule
         Property                               & Value                     & Ref.  \\
        \midrule
         $\alpha_{ICRS}$, J2000 {[}hh mm ss{]}  & 01:10:55.573              & 1     \\
         $\delta_{ICRS}$, J2000 {[}dd mm ss{]}  & +00:18:50.52              & 1     \\
         $\mu_{\alpha}$ {[}mas yr$^{-1}${]}     & $14.642\pm0.026$          & 1     \\
         $\mu_{\delta}$ {[}mas yr$^{-1}${]}     & $-8.103\pm0.019$          & 1     \\
         $\varpi$ {[}mas{]}                     & $2.0152\pm0.0230$         & 1     \\
         Distance {[}pc{]}                      & $486.6^{+6.7}_{-5.9}$     & 2     \\ 
        \midrule
         $G$ {[}mag{]}                          & $14.2645\pm0.0028$        & 1     \\
         $G_{BP}$ {[}mag{]}                     & $14.7271\pm0.0033$        & 1     \\
         $G_{RP}$ {[}mag{]}                     & $13.6425\pm0.0039$        & 1     \\
         $J$ {[}mag{]}                          & $12.897\pm0.026$          & 3     \\
         $H$ {[}mag{]}                          & $12.431\pm0.024$          & 3     \\
         $K_s$ {[}mag{]}                          & $12.321\pm0.024$          & 3     \\
         $u$ (AB) {[}mag{]}                     & $17.63\pm0.02$            & 4     \\
         $g$ (AB) {[}mag{]}                     & $16.096\pm0.008$          & 4     \\
         $r$ (AB) {[}mag{]}                     & $16.69\pm0.02$            & 4     \\
         $i$ (AB) {[}mag{]}                     & $14.123\pm0.005$          & 4     \\
         $z$ (AB) {[}mag{]}                     & $14.832\pm0.011$          & 4     \\
        \midrule
         $R_*$ [\rsun{}]                        & $0.830\pm0.022$           & 5     \\
         $M_*$ [$M_{\odot}$]                    & $0.85\pm0.02$             & 5     \\
         {[}Fe/H{]} [dex]                       & $+0.02^{+0.15}_{-0.14}$   & 6     \\
         log\,$g$ [log$_{10}$ cm\,s$^{-2}$]     & $4.56^{+0.03}_{-0.05}$    & 6     \\
         \teff{} [K]                            & $5060\pm50$               & 5     \\
         $f_{bol}$ [10$^{-11}$ erg s$^{-1}$]    & $5.288\pm0.130$           & 5     \\
         $m_{bol}$ [mag]                        & $14.194\pm0.027$          & 5     \\
         $M_{bol}$ [mag]                        & $5.716\pm0.036$           & 5     \\
         $L_{bol}$ [$L_{\odot}$]                & $0.4070\pm0.0137$         & 5     \\
         log($L/L_{bol}$) [dex]                 & $-0.3904\pm0.0146$        & 5     \\
         
        \bottomrule
    \end{tabular}
    \tablefoot{References:
    (1) Gaia EDR3 \citep{GaiaEDR3},
    (2) \citet{BailerJones21},
    (3) 2MASS \citep{Cutri03},
    (4) SDSS DR8,
    (5) this work,
    (6) StarHorse \citep{Anders19}.
    }
    \label{tab:Stellarprop}
\end{table}
\egroup

\section{Analysis}
\label{sec:analysis}

\subsection{Stellar Parameters} 

To independently estimate basic stellar parameters for \ac{etwo}, we use the Virtual Observatory SED Analyzer 6.0 \citep[VOSA;][]{Bayo08}\footnote{http://svo2.cab.inta-csic.es/theory/vosa/} to construct the star's Spectral Energy Distribution (SED) and fit synthetic stellar spectra. 
For reddening, we adopt the full Galactic dust column in the direction of the star from \citet{Schlafly11} ($E$($B$-$V$) = $0.0219\pm0.0003$) as the STILISM 3D reddening maps from \citep{Lallement18} show that the dust in the direction of the star is mostly confined to within $d < 335$ pc and the star lies at a Gaia EDR3-inferred distance of $d = 487\pm6$ pc \citep{BailerJones21}.
For dwarf stars of \teff\, $\simeq$ 5000\,K, a typical Galactic ratio of total to selection extinction value of $R_V$ (=$A_V$/$E$($B$-$V$)) is 3.2 \citep{McCall04}, with $\sigma_{R_V}$ $\simeq$ 0.18 
so we adopt an interstellar extinction $A_V$ = 0.070. 
Using the relation $A_{Ks}$ = 0.382\,$E$($B$-$V$) from  \citet{Bilir08}, we estimate the $K_s$ extinction to be $A_{Ks}$ = 0.0084. 
Photometry from several surveys was included in the SED fitting, incuding: {\it GALEX NUV} \citep{Bianchi11}, {\it Sloan} DR9 $ugriz$ \citep{AdelmanMcCarthy11}.

\begin{figure*}[ht]
    \centering
    \includegraphics[width=\linewidth]{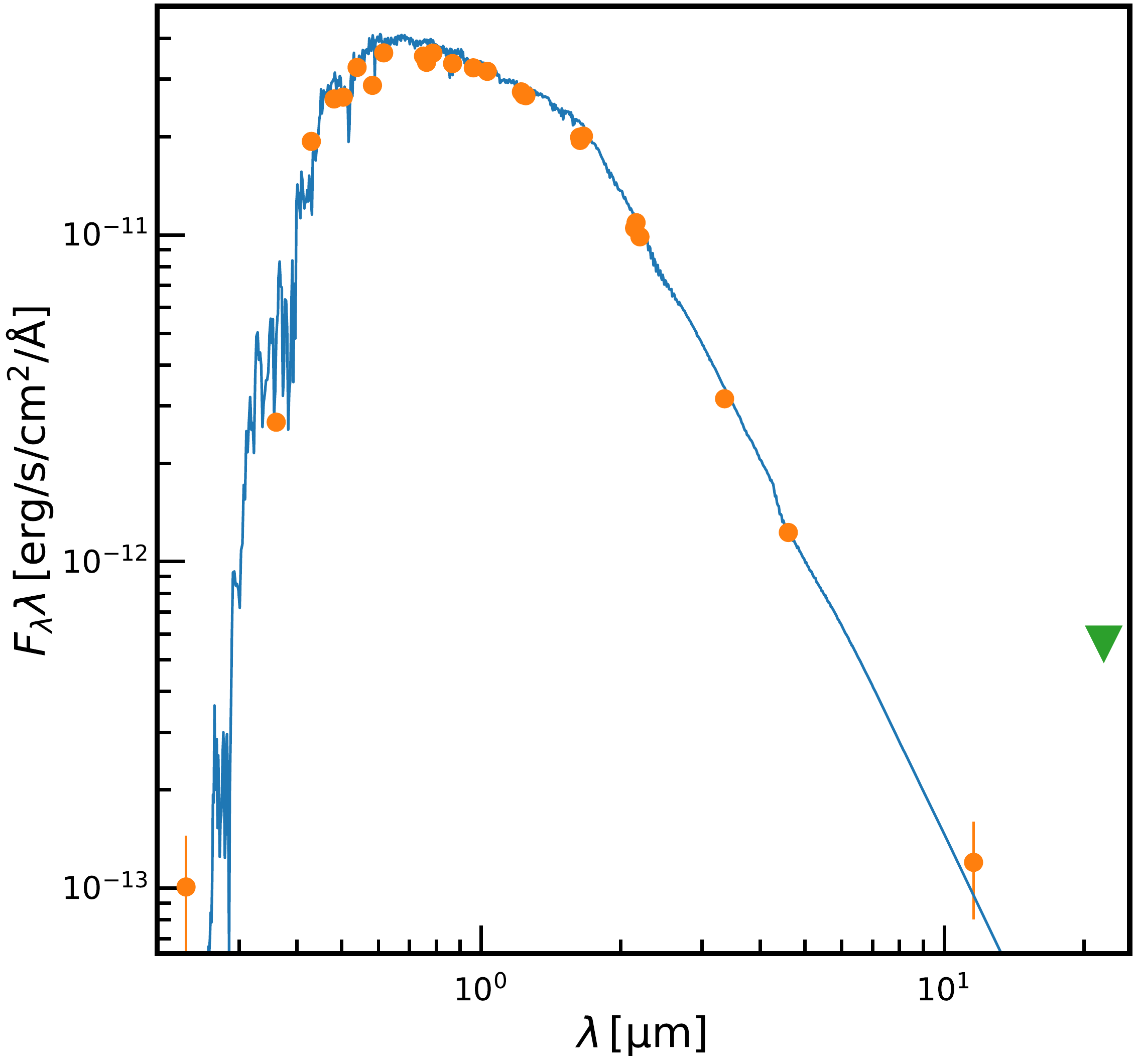}
    \centering
    \caption{The spectral energy distribution for EPIC~2202. The blue curve is a Kurucz best fit model. The orange points are the photometry as compiled from SIMBAD. The green triangle is an upper limit measurement from WISE at 22 microns.}
    \label{fig:SED}
\end{figure*}

For the fits of the photometry to the synthetic stellar spectra (see Figure~\ref{fig:SED}), we constrained the extinction $A_V$ $\in$ [0.06, 0.08], surface gravity \logg{} $\in$ [4, 5], and metallicity [M/H] $\in$ [-0.5, 0.5], as the star is clearly a dwarf and the photometric metallicities are approximately solar with [M/H] $= +0.02^{+0.15}_{-0.14}$ \citep{Anders19}.
For a grid of Kurucz ATLAS9 models \citep{Castelli03}, the best fit spectral template (fitting 27 photometric bands) was \teff{} = 5000\,K, [M/H] = +0.5, \logg{} = 4, $L = 0.407\pm0.010\, L_{\odot}$.
Varying whether the UV photometry ({\it GALEX NUV}) or reddest IR bands ({\it WISE W3 and W4}) were included or not, and examining the scatter in output parameters when varying metallicity, surface gravity, or extinction within the previous bounds, all had negligible effect on the resultant bolometric flux and luminosity. 
No ultraviolet ({\it NUV}) or infrared (e.g. {\it WISE W3 and W4}) excess was apparent.
From this SED analysis we adopt the following parameters:
\begin{itemize}
    \item bolometric flux $f_{bol}$ = 5.288($\pm$0.130) $\times 10^{-11}$ erg s$^{-1}$ cm$^{-2}$,
    \item apparent bolometric magnitude $m_{bol}$ = $14.194\pm0.027$ (IAU 2015 scale),
    \item absolute bolometric magnitude $M_{bol}$ = $5.716\pm0.036$ (IAU 2015 scale),
    \item bolometric luminosity $L = 0.4070\pm0.0137$ $L_{\odot}$,
    \item log($L/L_{bol}$) = $-0.3904\pm0.0146$ dex.
\end{itemize}

There are several other published \teff{} estimates for \ac{etwo}, listed in Table \ref{tab:teffs}. 
The \teff{} estimates are in reasonable statistical agreement with one another.
The \citet{Anders19} value is somewhat high, likely due to adopting an excessive extinction ($A_G$ = 0.20) which does not appear to be supported by the extinction maps \citep{Schlafly11, Lallement18}.
Omitting the \citet{Anders19} value, we adopt the unweighed mean \teff{} of the remaining values, and adopt the standard deviation as a conservative estimate of the uncertainty: \teff{} = $5060\pm50$ K (1.0\%\, uncertainty).
This \teff{} is typical for a solar composition K2V star \citep{Pecaut13}\footnote{\url{http://www.pas.rochester.edu/~emamajek/EEM_dwarf_UBVIJHK_colors_Teff.txt}}.
Combining the adopted \teff{} estimate with the previously estimated luminosity from our SED analysis, we estimate the radius of \ac{etwo} to be 
$R$ = $0.830\pm0.022$ $R_{\odot}$ 
(IAU nominal solar radius) or $577\,600\pm15\,100$ km (2.6\% uncertainty).

\begin{table}
    \caption{Estimated Effective Temperatures for \ac{etwo}.}
    \label{tab:teffs}
    \centering
    \begin{tabular}{ll}
        \toprule
        \toprule
         \teff{} [K]            & Reference                 \\
        \midrule                        
         $4997^{+97}_{-46}$     & \citet{GaiaDR2}           \\
         $5000$                 & this work                 \\
         $5064^{+102}_{-81}$    & \citet{Huber16}           \\
         $5090\pm46$            & \citet{Bai19}             \\
         $5095\pm122$           & \citet{Stassun19}         \\
         $5105\pm138$           & \citet{HardegreeUllman20} \\
         $5206^{+156}_{-104}$   & \citet{Anders19}          \\
        \midrule
         {\bf $5060\pm50$}      & {\bf adopted}             \\
        \bottomrule
    \end{tabular}
\end{table}

Only two mass estimates have been published since the availability of parallax data for \ac{etwo} by Gaia DR2.
\citet{Anders19} (StarHorse) estimates the mass to be $0.827^{+0.057}_{-0.031}$ \Msun{}, and
the TIC version 9 \citep{Stassun19}
estimates the mass to be 0.850 \Msun{} (no uncertainties). 
\citet{Huber16} estimated the mass to be $0.816^{+0.050}_{-0.070}$ \Msun{}, however this was without the benefit of a trigonmetric parallax. 
Using the most recent mass-luminosity calibration for binary stars with dynamical masses from \citet{Eker18}, our luminosity estimate (\logl\, = -0.390) translates to a mass estimate of 0.858 \Msun{}.
Based on these three estimates \citep[][ and this work]{Stassun19,Anders19}, we adopt a stellar mass of $0.85\pm0.02$ \Msun{}. 


\begin{figure*}[t]
    \centering
    \includegraphics[width=\linewidth]{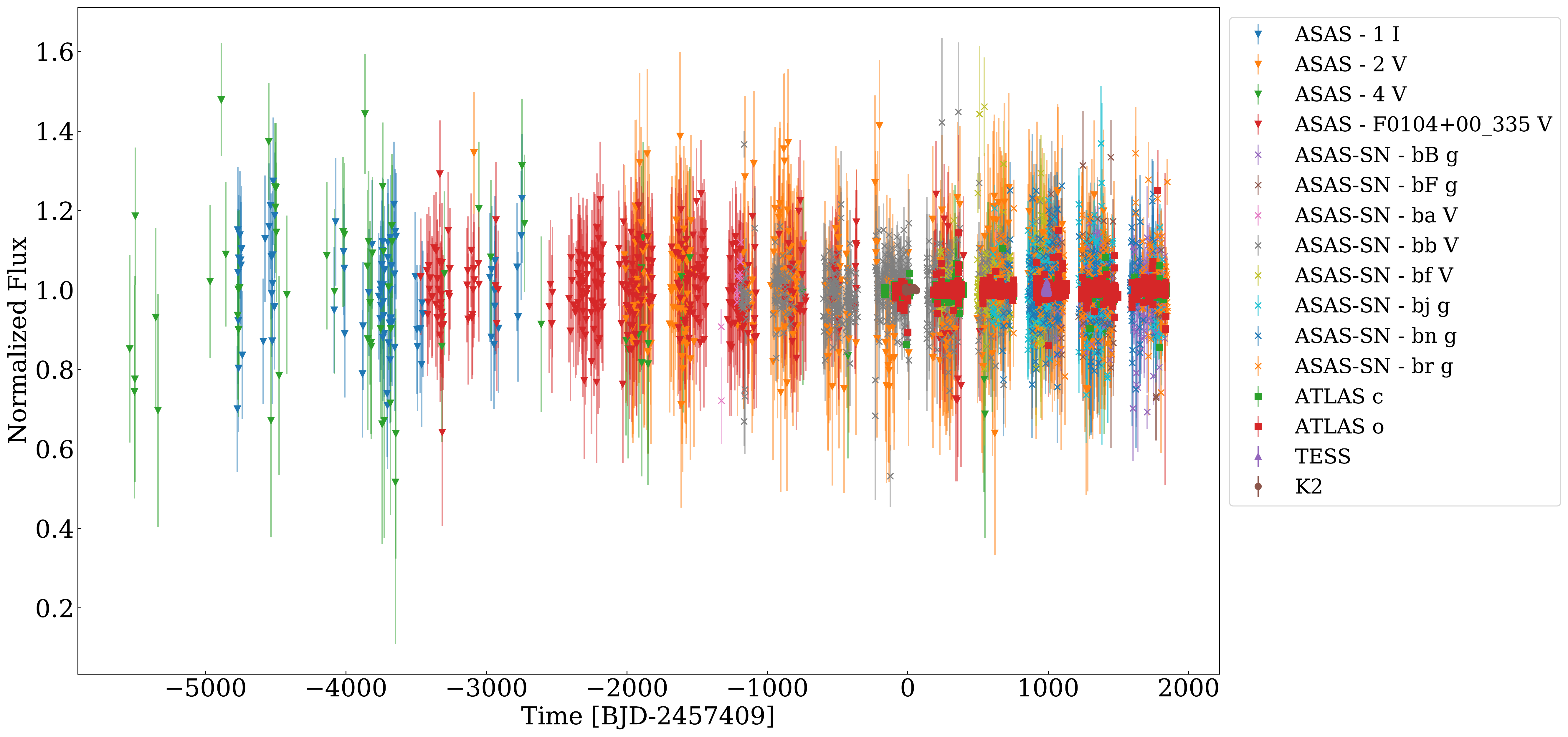}
    \caption{Time-series photometry of \ac{etwo} from several ground-based surveys as well as the \textit{K2} and TESS data.
      Photometry with a flux value greater than 150\% or less than 10\% was excluded, along with flux errors error greater than 100\%.}
    \label{fig:groundbased}
    \vspace{5pt}
\end{figure*}

\subsection{Eclipse Modeling}\label{sec:dip} 

The eclipse observed in \textit{K2} photometry is most likely due to an occulter transiting \ac{etwo}.
The asymmetry and depth of the eclipse, as well as the short duration, make it unlikely for the transiting object to be a sphericlly symmetric object.
We assume that the eclipse is due to a tilted and inclined disk gravitationally bound to an unseen secondary companion that creates an elliptical occulter due to its projected geometry. 
We model the light curve with two models - a hard edged disk, and a two ring disk consisting of an inner opaque disk and an outer, partially transmissive ring.

\subsubsection{Model Parameters}

The modeling is performed using a modified version of the \texttt{pyPplusS} package that models the transit of an oblate exoplanet or an exoplanet with rings by calculating the area hidden by the occulter across the limb-darkened disk of the star \citep{rein_vizier_2019}.
We use this to model the light curve based on the position and geometry of the disk with respect to the (limb-darkened) host star.
The code works the spatial domain (in units of stellar radii) and must be converted to the temporal domain, by introducing a transverse velocity $v_t$ and fitting for $\delta t$ w.r.t. $BJD =$ 2457409, to produce a light curve.

To calculate a lower bound on $v_t$ we use the method described in \cite{van_werkhoven_analysis_2014}.
This method requires the linear limb-darkening parameter, $u$, of the star, which has been determined using the \texttt{jkltd} program written by \cite{southworth_jktld_2015}.
This program uses the surface gravity, $\log{\mathrm{g}}$, the effective temperature, \teff{}, and the metallicity, [Fe/H], of the star to linearly interpolate the tables from \cite{sing_stellar_2010} to calculate $u$ for the \textit{Kepler} bandpass.
These values are derived by \cite{huber_k2_2016} for 138,600 \textit{K2} targets in campaigns 1-8, using proper motions, colors, parallaxes, spectroscopy and stellar population models.
For \ac{etwo} we have $\log{\mathrm{g}} = 4.591$, \teff{} = 5064 K and [Fe/H] = -0.112, which results in a value of $u = 0.6681$.

Using $u$, the steepest time gradient of the light curve, $\dot{L}$, and taking $R = R_*$, we can get a lower limit on $v_t$ following the method of \cite{van_werkhoven_analysis_2014}:
\begin{ceqn}
    \begin{align}
        \centering
        v_t = \dot{L}R\pi \left( \frac{2u-6}{12-12u+3\pi u}\right).
    \end{align}
\end{ceqn}
For our obtained value of $u$ and for $\dot{L}$ = 3 $L_*$ day$^{-1}$ we obtain a lower limit of $v_t$ = 4.3 $R_*$ day$^{-1}$.\\

The free parameters for the model are the radius of the disk, $R_d$, the impact parameter (perpendicular distance between the center of the star and the orbital path), $b$, the inclination of the disk, $i$, the tilt of the disk (the angle w.r.t the orbital path), $\phi$, the transverse velocity of the disk, $v_t$, the time shift w.r.t the time of closest approach, $\delta t$ and the opacity of the disk $\tau$.

We use \texttt{emcee} \citep{foreman-mackey2013} to explore these parameters and use Markov Chain Monte Carlo (MCMC) methods to determine the best fit of the models.
To properly explore this parameter space physical bounds on these parameters must be applied.
These are as follows: an upper limit for $R_d$ of ten times the radius of the star, $R_*$,
the upper and lower limit on $b$ are such that the disk must transit the star,
the bounds on $i$ are from 0$^\circ$ (face-on disk) to 90$^\circ$ (edge-on disk), 
$\phi$ has been limited from 0$^\circ$ to 90$^\circ$ because of reflection symmetries induced by the combination of $b$ and $\phi$, 
and a range of $\delta t$ between -10 and 10 days.
We take $v_t$ between 4.3 $R_*$ day$^{-1}$ and 20 $R_*$ day$^{-1}$, where the upper bound has been deemed large enough and the lower bound has been calculated above.
The opacity of the disk, $\tau$, can generally be bound between 0 and 1, but here we fix $\tau$ at 1, to get the smallest possible disk. 

\bgroup
\begin{table}[htb!]
    \centering
    \caption{Parameter Bounds for MCMC Optimisation.}
    \begin{tabular}{lcc}
        \toprule
        \toprule
         Parameter                      & Lower Bound   & Upper Bound   \\ 
        \midrule
         $R_d$ {[}$R_*${]}              & 0             & 10            \\
         $t_e$ {[}$R_*${]}              & 0             & 10            \\
         $b$ {[}$R_*${]}                & -10           & 10            \\
         $i$ {[}$^\circ${]}             & 0             & 90            \\
         $\phi$ {[}$^\circ${]}          & 0             & 90            \\
         $v_t$ {[}$R_*$ day$^{-1}${]}   & 4.3           & 20            \\
         $\delta t$ {[}day{]}           & -10           & 10            \\
         $\tau$ {[}-{]}                 & 1             & 1             \\
         $\tau_e$ {[}-{]}                 & 0             & 1             \\
        \bottomrule
    \end{tabular}
    \label{tab:MCMCpar}
\end{table}
\egroup

\subsubsection{Hard Edged Disk}

The first model we explore is a hard edged disk.
We fix the opacity such that the disk is completely opaque, since this gives the smallest possible disk diameter for a given light curve gradient.
The smallest physically plausible disk model then gives a lower bound on the mass of the secondary companion through the size of the companion's Hill sphere.
We are interested in the smallest possible disks as these are deemed more likely to exist due to stability of the disk, and orbital velocity considerations.

To determine a suitable starting point of the fit we set a Gaussian prior bound by the limits in Table \ref{tab:MCMCpar} and described in the previous section.
We ran this for 600 links with 1000 walkers.
We used a local minimum from this parameter investigation to set the initial starting point for the walkers of the extended optimisation.
This starting point was expanded with Gaussian priors to produce the initial parameter spread summarised in Table \ref{tab:MCMCpriors}.
The extended optimisation was performed with 1000 walkers for 700 links using the priors and parameters described.

\bgroup
\begin{table}[htb!]
    \centering
    \caption{Gaussian Priors for MCMC Optimisation.}
    \begin{tabular}{lcccc}
        \toprule
        \toprule
         Parameter                      & \multicolumn{2}{c}{Hard Edged Disk}   & \multicolumn{2}{c}{Soft Edged Disk}   \\
                                        & Median        & Spread                & Median        & Spread                \\ 
        \midrule
         $R_d$ {[}$R_*${]}              & 0.9           & 0.1                   & 1.4           & 0.1                   \\
         $t_e$ {[}$R_*${]}              & -             & -                     & 0.1           & 0.1                   \\
         $b$ {[}$R_*${]}                & 0.3           & 0.1                   & 0.7           & 0.1                   \\
         $i$ {[}$^\circ${]}             & 72            & 6                     & 77            & 6                     \\
         $\phi$ {[}$^\circ${]}         & 49            & 6                     & 37            & 6                     \\
         $v_t$ {[}$R_*$ day$^{-1}${]}   & 10.0          & 0.5                   & 11.6          & 0.1                   \\
         $\delta t$ {[}day{]}           & -0.04         & 0.10                  & -0.04         & 0.01                  \\
         $\tau$ {[}-{]}                 & -             & -                     & -             & -                     \\
         $\tau_e$ {[}-{]}              & -             & -                     & 0.1           & 0.1                   \\ 
        \bottomrule
    \end{tabular}
    \label{tab:MCMCpriors}
\end{table}

\begin{figure}[ht]
    \hspace{-13pt}
    \begin{subfigure}{.55\textwidth}
      \centering
      \includegraphics[width=\linewidth]{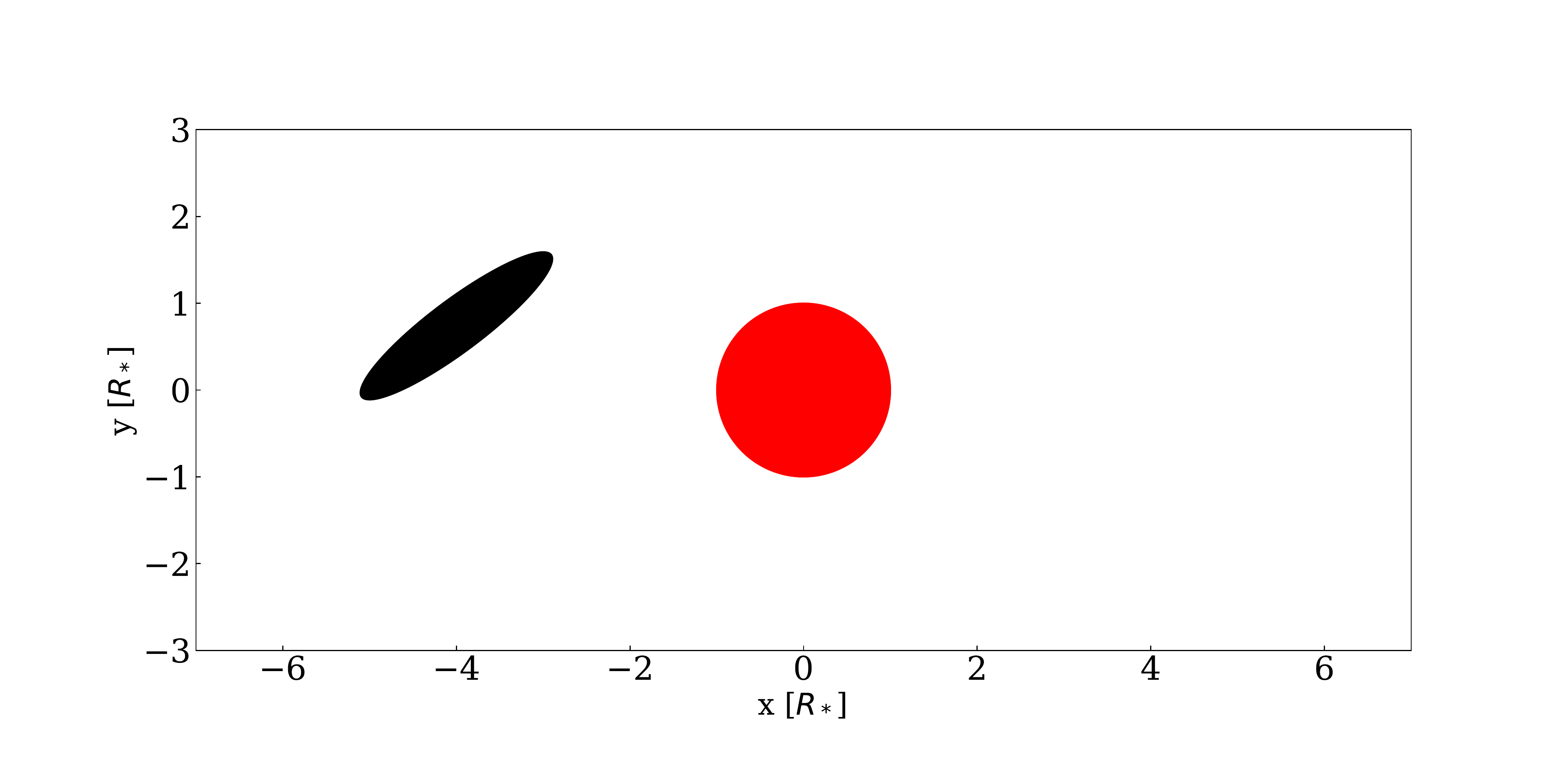}
      \label{fig:modeldepiction}
        \vspace{-30pt}
    \end{subfigure}
    \vspace{-20pt}
    \begin{subfigure}{.49\textwidth}
      \centering
      \includegraphics[width=\linewidth]{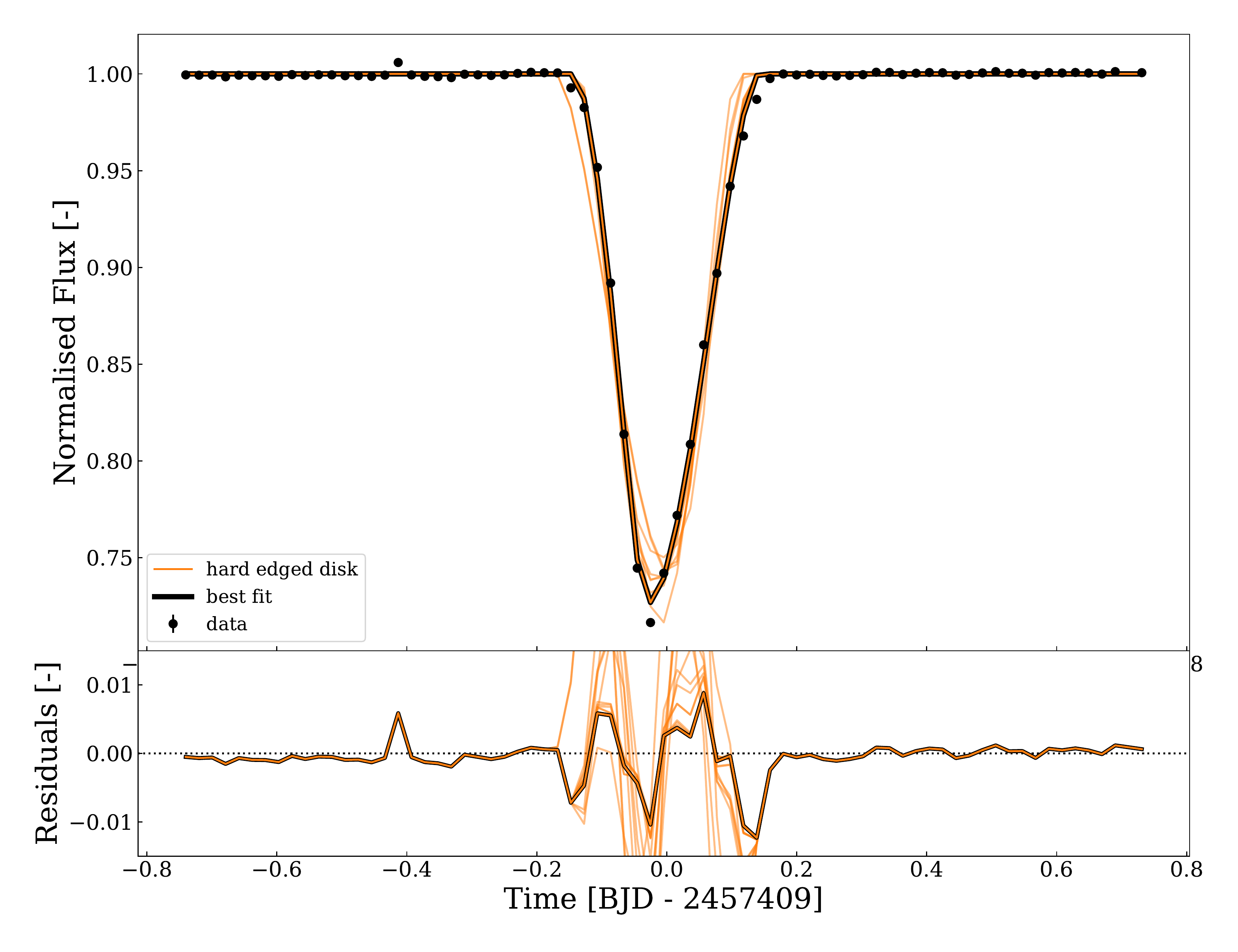} 
      \label{fig:hardedgefit}
    \end{subfigure}
    \caption{\textit{Upper}: depiction of the best fit result for the hard edged disk. \textit{Lower}: hard edged disk model fits of 300 random walkers with a burn-in of 500 links.}
    \label{fig:hardedgemodel}
\end{figure}

The results of the MCMC optimisation model fit and a physical depiction of the model are shown in Figure \ref{fig:hardedgemodel} and summarised in Table \ref{tab:modelparameters}. 
Note that the errors displayed in this table are on the MCMC distribution, and thus do not include any systematic errors in the photometry due to unmodeled astronomical noise sources, such as stellar variation.

The residuals of the fit are at around the 1 percent level. 
However, it is clear from the systematic deviations of the residuals that the modeled eclipse is not wide enough at the start and end of the eclipse and not deep enough at maximum occultation to follow the measured light curve.
This pattern suggests that the disk is not quite large enough and prompts an extension of the hard edged model.

\subsection{Soft Edged Disk}

Based on the one percent residuals that we see in the hard edged model, we added an outer ring with variable thickness, $t_e$, and opacity, $\tau_e$, and centre the prior of the soft edged disk model on the best fit of the hard edged disk model.
We set the bounds of $\tau_e$ to be between 0 and 1, and those of $t_e$ to be the same as $R_d$. 
See also Table \ref{tab:MCMCpar}.
The addition of a soft edge would increase the width and depth of the eclipse, but the opacity must be less than 1, otherwise this solution would have been found by the hard edged disk optimisation.
We initialised 1000 walkers for 1000 links for the soft edged disk model, using the Gaussian priors summarised in Table \ref{tab:MCMCpriors}. 
The results of the MCMC optimisation model fit and a physical depiction of the model are shown in Figure \ref{fig:softedgemodel} and summarised in Table \ref{tab:modelparameters}. 

\bgroup
\begin{table}[htb!]
    \begin{center}
        \caption{Results of the MCMC Optimisation.}
        \begin{tabular}{lcc}
            \toprule
            \toprule
             Parameters                     & Hard Edged Disk           & Soft Edged Disk                       \\ 
            \midrule
             $R_d$ {[}$R_*${]}              & 1.374 $\pm$ 0.002         & 1.163 $\pm$ 0.005                     \\
             $t_e$ {[}$R_*${]}              & -                         & 0.317 $\pm$ 0.005                     \\
             $b$ {[}$R_*${]}                & 0.7397 $\pm$ 0.0010       & 0.7537 $\pm$ 0.0011                   \\
             $i$ {[}$^\circ${]}             & 77.01 $\pm$ 0.03          & 75.94 $\pm$ 0.03                      \\
             $\phi$ {[}$^\circ${]}          & 36.81 $\pm$ 0.05          & 38.04 $\pm$ 0.07                      \\
             $v_t$ {[}$R_*$ day$^{-1}${]}   & 11.589 $\pm$ 0.007        & 11.501 $^{+0.007}_{-0.006}$           \\
             $v_t$ {[}km s$^{-1}${]}        & 77.45 $\pm$ 0.05          & 76.86 $\pm$ 0.04                      \\
             $\delta t$ {[}day{]}           & -0.04083 $\pm$ 0.00008    & -0.04040 $^{+0.00009}_{-0.00008}$     \\
             $\tau$ {[}-{]}                 & 1                         & 1                                     \\
             $\tau_e$ {[}-{]}               & -                         & 0.499 $^{+0.008}_{-0.011}$            \\ 
            \bottomrule
        \end{tabular}
        \label{tab:modelparameters}
    \end{center}
    \begin{tablenotes}
        \item  Note: transverse velocity is converted from $R_*$ day$^{-1}$ to km s$^{-1}$ using $R_*$ = 0.83 \rsun.
    \end{tablenotes}
\end{table}
\egroup

\begin{figure}[ht]
    \hspace{-13pt}
    \begin{subfigure}{.55\textwidth}
      \centering
      \includegraphics[width=\linewidth]{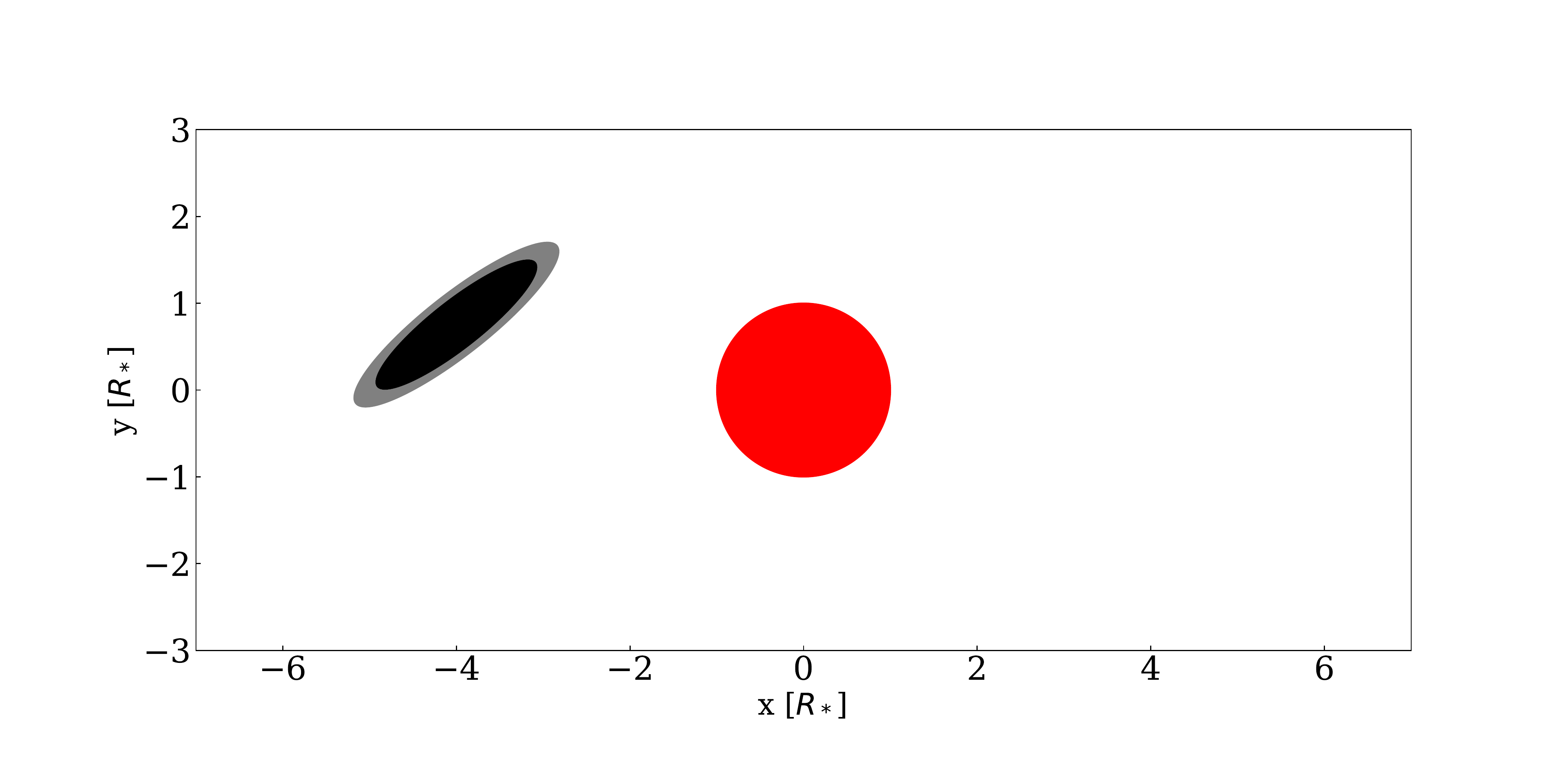}
      \label{fig:modeldepiction_soft}
    \vspace{-30pt}
    \end{subfigure}
    \vspace{-20pt}
    \begin{subfigure}{.49\textwidth}
      \centering
      \includegraphics[width=\linewidth]{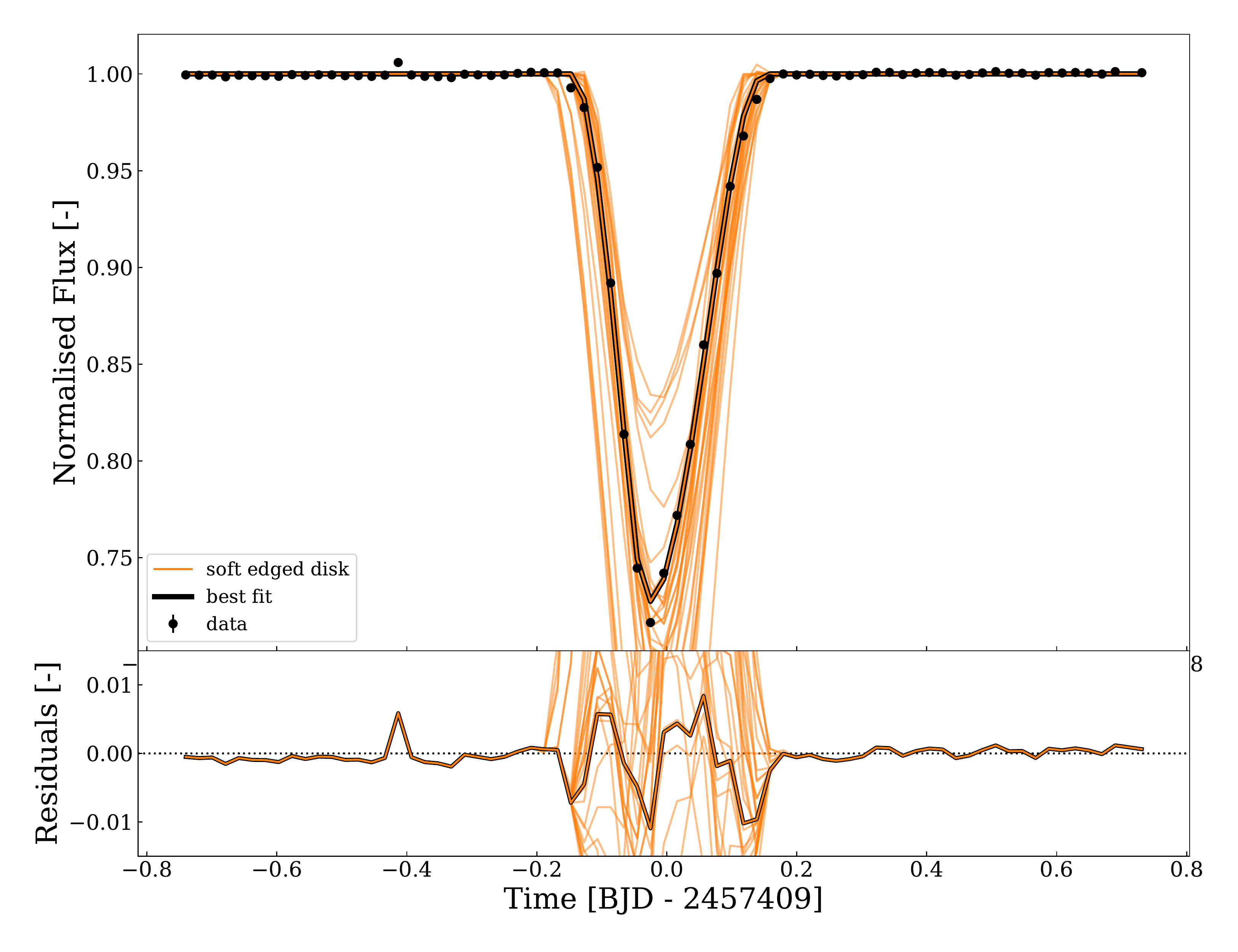} 
      \label{fig:softedgefit}
    \end{subfigure}
    \caption{\textit{Upper}: depiction of the best fit result for the hard edged disk. \textit{Lower}: soft edged disk model fits of 300 random walkers with a burn-in of 550 links.}
    \label{fig:softedgemodel}
\end{figure}

\begin{figure}[htb!]
    \centering
    \includegraphics[scale=0.27]{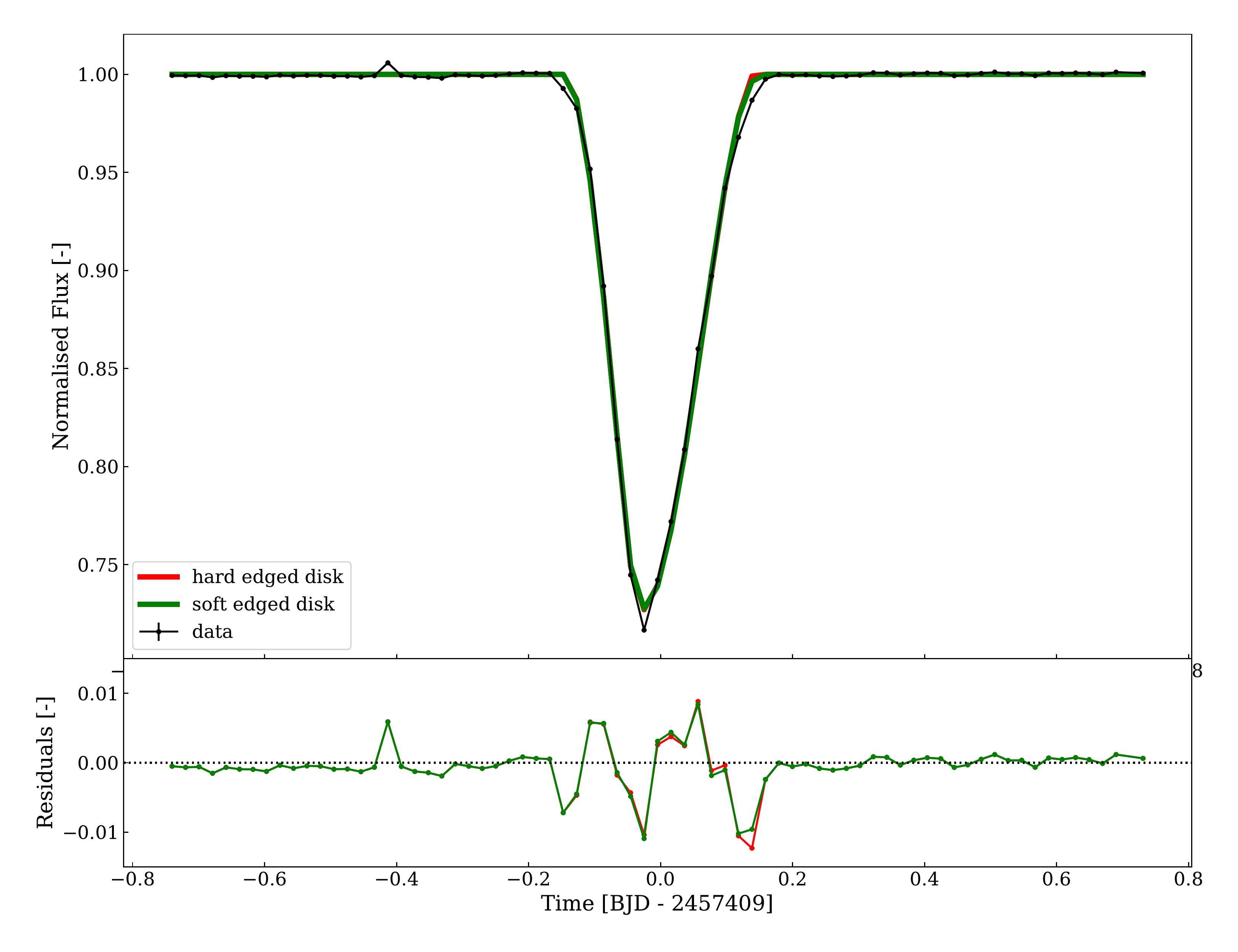}
    \caption{ A comparison of the hard edged model and the soft edged model with residuals.}
    \label{fig:comparisonplot}
    \vspace{5pt}
\end{figure}

\subsubsection{Comparison of the Two Models}

The residuals from both models are similar in amplitude and shape, as can be seen in Figure \ref{fig:comparisonplot}, which overlays the best fits.
The soft edged model does give a slightly lower $\chi^2$ value (65,955 for 72 photometric points) compared to the hard edged model (70,729 for 72 photometric points), which implies that the soft edged model is a slightly better fit to the data.
However, it is also clear that the addition of a soft edge to the hard edge model was not enough to reduce the observed residuals and thus does not explain their origin.

\subsection{Period Folding}
\label{sec:period}

To search for a plausible orbital period of an occulter orbiting \ac{etwo}, we supplemented the data from \textit{K2} with data from TESS and several ground-based surveys, as can be seen in Figure \ref{fig:groundbased}.

We folded all the data over a large (60-1800 days) high temporal resolution (step size of 0.001 days) period grid and examined the folded light curve in the region where the \textit{K2} eclipse occurs. 
The routine flags periods where at least three photometric points lie within three times the photometric error of that point from the linearly interpolated \textit{K2} data.

The best period candidates were determined by comparing the $\chi^2$ value of folded data w.r.t. the \textit{K2} data ($\chi^2_{model}$) to the $\chi^2$ value of the folded data w.r.t. a flat line ($\chi^2_{flat}$, i.e. a no eclipse model).
The lower the ratio $\chi^2\mathrm{-ratio} = \chi^2_{flat}/\chi^2_{model}$, the more the folded data follows a no-eclipse scenario.
The higher the $\chi^2\mathrm{-ratio}$, the more likely the folded data follows an eclipse scenario.
A further criterion was that at least one point of the folded data lies in the phase interval where the flux of the \textit{K2} data is below a flux of 85\%.
This is to ensure that the periods investigated actually contain data in the deepest part of the eclipse.

Of all the periods investigated, 13 out of 15 periods with a $\chi^2\mathrm{-ratio} >$ 400 suggest a fundamental period of 290.230 days, and 2 out of 15 periods suggest a fundamental period of 235.587 days (see Table \ref{tab:periodfolding} for more details).
For a period of 290.230 days, photometry with a $\chi^2\mathrm{-ratio} >$ 400 was found for 2 and 4 times the period, but not for 3 times the period due to incomplete photometric coverage.
For a period 235.587 days there was no reliable photometry for a multiple of the period, due to incomplete coverage or due to the absence of points in the deepest part of the eclipse (with flux values below 85\%).

Note that any of the periods where no photometry was folded into the eclipse can not be ruled out by this period folding analysis.
Therefore, there are many more possible periods for \ac{etwo} that can not be investigated with the available data.


\bgroup

\begin{table}[htb!]
    \centering
    \caption{Reasonable Fundamental Periods ($\chi^2\mathrm{-ratio} >$ 400).}
    \begin{tabular}{lcc}
        \toprule
        \toprule
         Period  & $n_{phot}$        &  $\chi^2\mathrm{-ratio}$          \\
         (day)   & (within eclipse)  & ($\chi^2_{flat}/\chi^2_{model}$)  \\
        \midrule
         290.230 & 4                 & 423.564   \\
         235.587 & 3                 & 410.788   \\ 
        \bottomrule
    \end{tabular}
    \label{tab:periodfolding}
\end{table}
\egroup

\begin{figure}[htb]
    \centering
    \includegraphics[width=\linewidth]{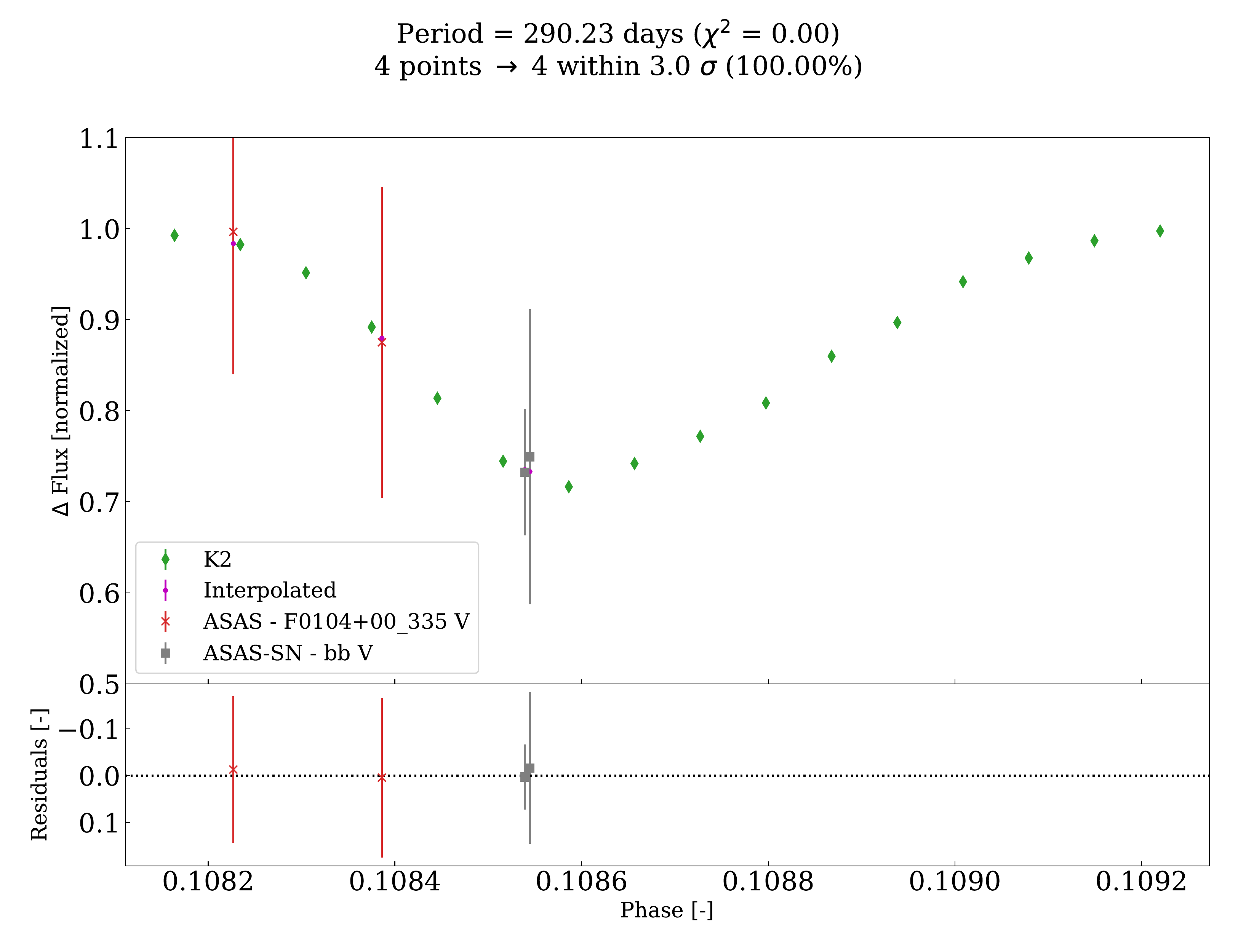}
    \caption{Phased Light Curve centred around \textit{K2} eclipse with a fold period of 290.230 days.}
    \label{fig:periodfoldingbest}
    \vspace{5pt}
\end{figure}

Considering the most likely shorter orbital periods (see Table \ref{tab:periodfolding}), we can make predictions for the upcoming eclipses of \ac{etwo}.
The dates, times and relation to the observed eclipse for the upcoming four eclipses for each period are listed in Table \ref{tab:predictions}.

\bgroup

\begin{table}[ht]
    \centering
    \caption{Next Transit Predictions.}
    \begin{tabular}{lccc}
        \toprule
        \toprule
         Julian Date     & Calendar Date & UT        & \begin{tabular}[c]{@{}c@{}} Nr. of periods \\ after \textit{K2} eclipse  \end{tabular}     \\ 
        \midrule
        \rowcolor{lichtgrijs}
         2459529.083    & 10-11-2021    & 18:47:31  & 9     \\
         2459730.640    & 31-05-2022    & 08:09:36  & 8     \\
        \rowcolor{lichtgrijs}
         2459764.670    & 04-07-2022    & 08:52:48  & 10    \\
        \rowcolor{lichtgrijs}
         2460000.257    & 24-02-2023    & 22:58:05  & 11    \\
         2460020.870    & 17-03-2023    & 13:40:48  & 9     \\
        \rowcolor{lichtgrijs}
         2460235.844    & 18-10-2023    & 13:03:22  & 12    \\
         2460311.100    & 01-01-2024    & 19:11:00  & 10    \\ 
        \bottomrule
        \end{tabular}
    \label{tab:predictions}
    \begin{tablenotes}
        \item grey rows correspond to $P_{orb} = 235.587$ days.
        \item white rows correspond to $P_{orb} = 290.230$ days.
      \end{tablenotes}
\end{table}
\egroup

\subsection{Orbital Parameters}
\label{sec:orbit}

We have obtained two transverse velocities from our hard edged and soft edged model, 11.6 and 11.5 $R_*$ day$^{-1}$ respectively, and taking  $R_*$ = 0.83 \rsun{} this corresponds to 77.4 and 76.9 km s$^{-1}$. 

The eclipse observed by \textit{K2} does not repeat within the observation window of K2, which gives us a lower limit on the orbital period, $P_{orb}$.
$P_{orb}$ must be larger than the longest time within the \textit{K2} data where no eclipse appears, which in this case is 60 days.

If we take the modelled transverse velocity, $v_t$, and calculate the period of a circular orbit, it results in $P_{orb} \approx 25$ days, which is below the minimum period limit set by the rest of the K2 data, meaning that a circular orbit with this velocity is ruled out.
We thus explore eccentric orbits, where we assume that $v_t = v_{peri}$, the velocity at periastron, and create a grid consisting $P_{orb}$, and the mass of the companion, $M_p$.
We follow the analysis as performed by \citet{van_dam_asymmetric_2020} in Section 4.4, which we describe shortly here.
Using Kepler's Third Law we use $P_{orb}$ to obtain the semi-major axis, $a$ - when combined with $v_{peri}$ and $M_p$, these then determine the eccentricity, $e$, using the vis-viva equation. 
Finally as a stability criterion, we state that $M_p$ must be large enough that the Hill radius, $R_{Hill}$, of the companion can support the circumsecondary disk ($R_{disk} < 0.3 \, R_{Hill}$).
We use the same $M_p$ bounds from 0 - 80 $M_{\mathrm{Jup}}$, with the upper limit chosen as an inclusive H-burning limit \citep{saumon_evolution_2008} and a more conservative limit of 73-74 $M_{\mathrm{Jup}}$ \citep{baraffe_new_2015,forbes_existence_2019}. 
For $P_{orb}$ we set limits of the grid from 60 - 1800 days, set the upper limit is considered arbitrarily large and is related to the fact that the available photometry does not extend further away from the observed eclipse. 
The results for the hard edged and soft edged disk models are indistinguishable due to their similar $v_t$.
The parameter space investigation is depicted in Figure \ref{fig:orbitalanalysis1800} and the orbital parameter ranges at the most likely periods (from Table \ref{tab:periodfolding}) are presented in Table \ref{tab:orbitalanalysis}.

\bgroup

\begin{table}[htb!]
    \centering
    \caption{Orbital Parameters for Most Likely Periods.}
    \begin{tabular}{lcc}
        \toprule
        \toprule
         Parameter           & $P_{orb} = 290$ days  & $P_{orb} = 235$ days  \\ 
        \midrule
         $r_{ap}$ {[}AU{]}   & 1.40 - 1.44           & 1.20 - 1.21           \\ 
         $r_{peri}$ {[}AU{]} & 0.22 - 0.24           &  0.22 - 0.23          \\
         $r_H$ {[}AU{]}      & 0.02 - 0.07           & 0.02 - 0.07           \\ 
         $e$ {[}-{]}         & 0.72 - 0.73           &  0.67 - 0.69          \\  
        \bottomrule
    \end{tabular}
    \label{tab:orbitalanalysis}
\end{table}
\egroup

\begin{figure*}[ht]
    \begin{center}
        \begin{subfigure}{.4\textwidth}
          \centering
          \includegraphics[width=\linewidth]{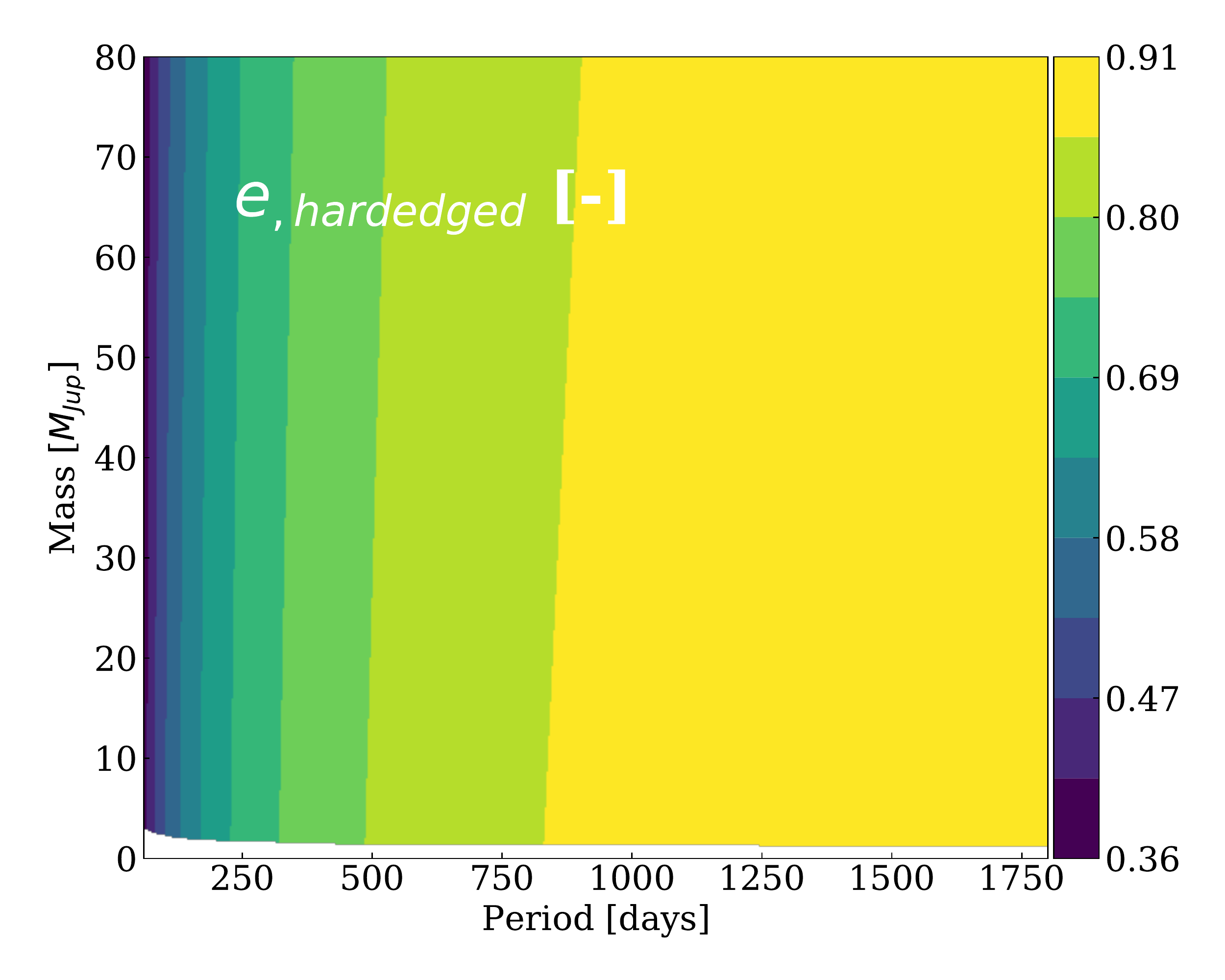} 
          \label{fig:ehardedged1800}
        \end{subfigure}
        \begin{subfigure}{.4\textwidth}
          \centering
          \includegraphics[width=\linewidth]{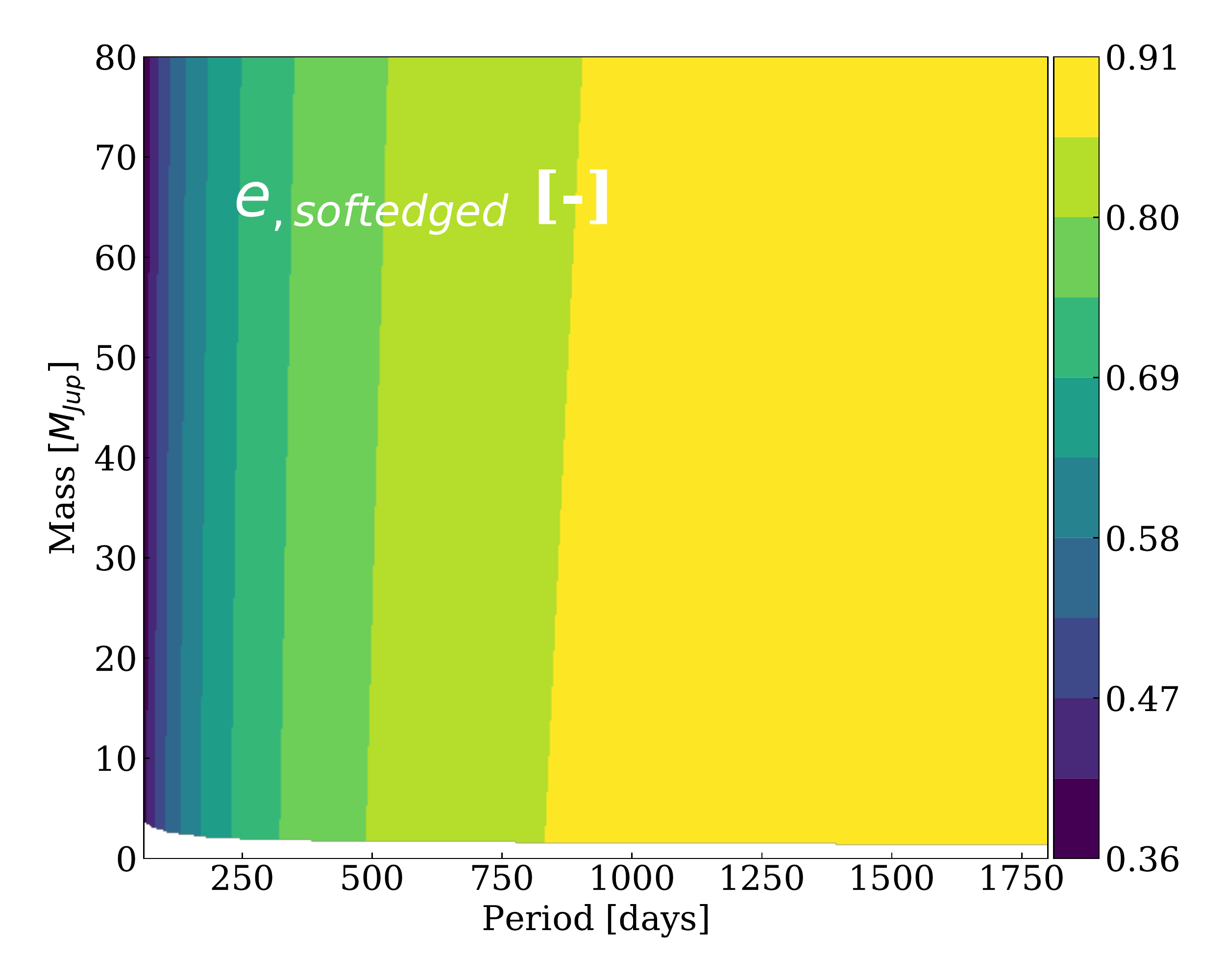}  
          \label{fig:esoftedged1800}
        \end{subfigure}
        \begin{subfigure}{.4\textwidth}
          \centering
          \includegraphics[width=\linewidth]{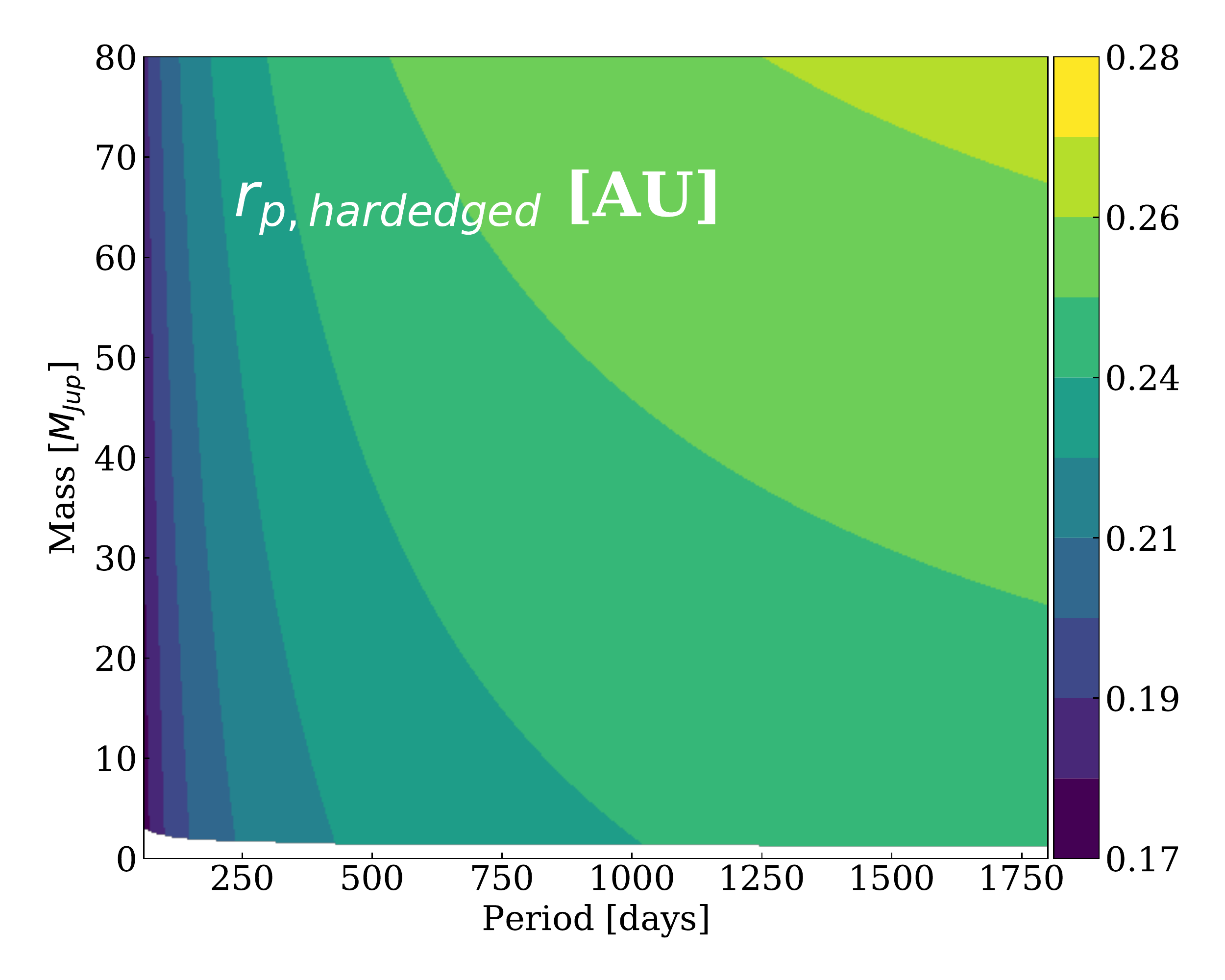} 
          \label{fig:raphardedged1800}
        \end{subfigure}
        \begin{subfigure}{.4\textwidth}
          \centering
          \includegraphics[width=\linewidth]{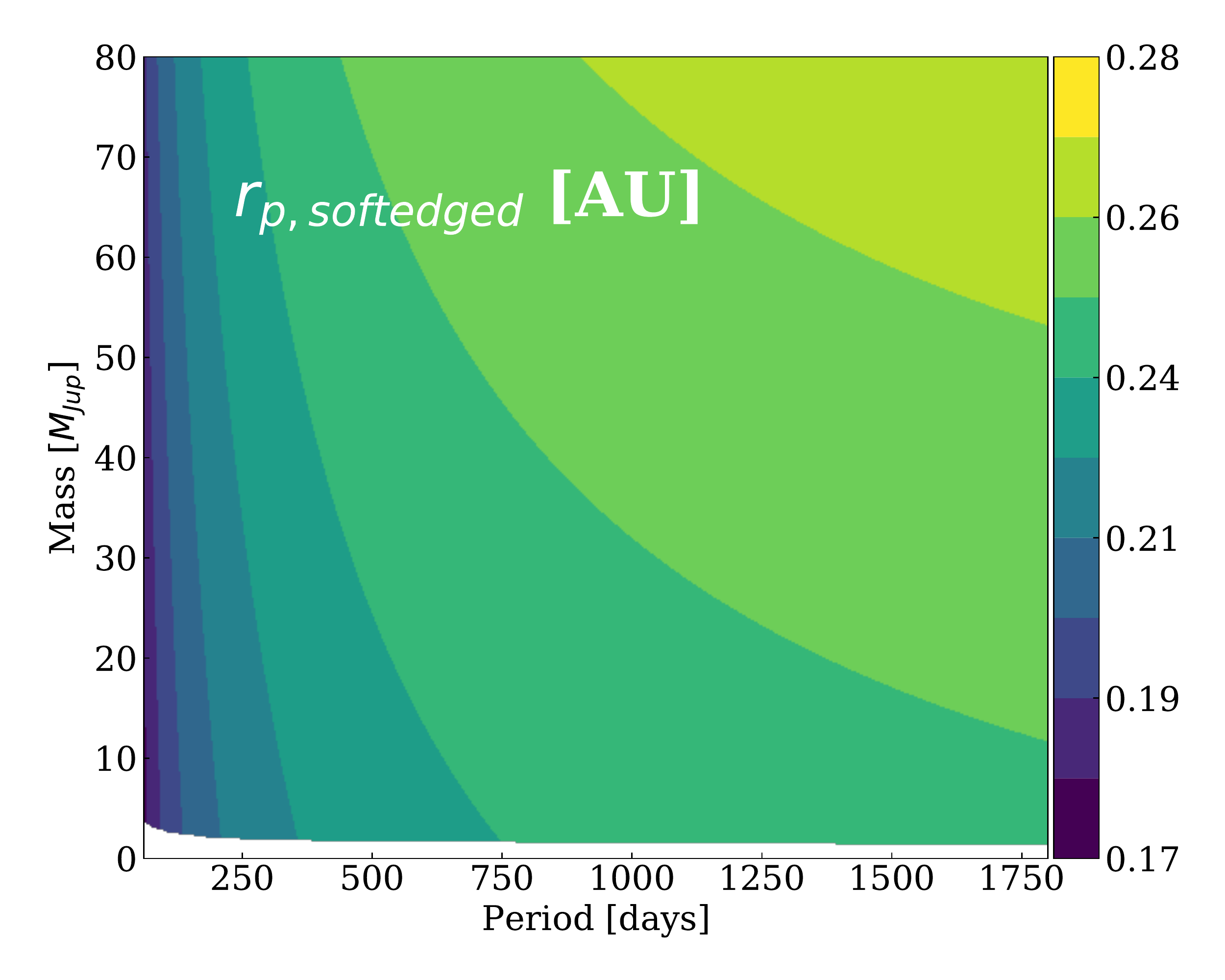}  
          \label{fig:rapsoftedged1800}
        \end{subfigure}
        \begin{subfigure}{.4\textwidth}
          \centering
          \includegraphics[width=\linewidth]{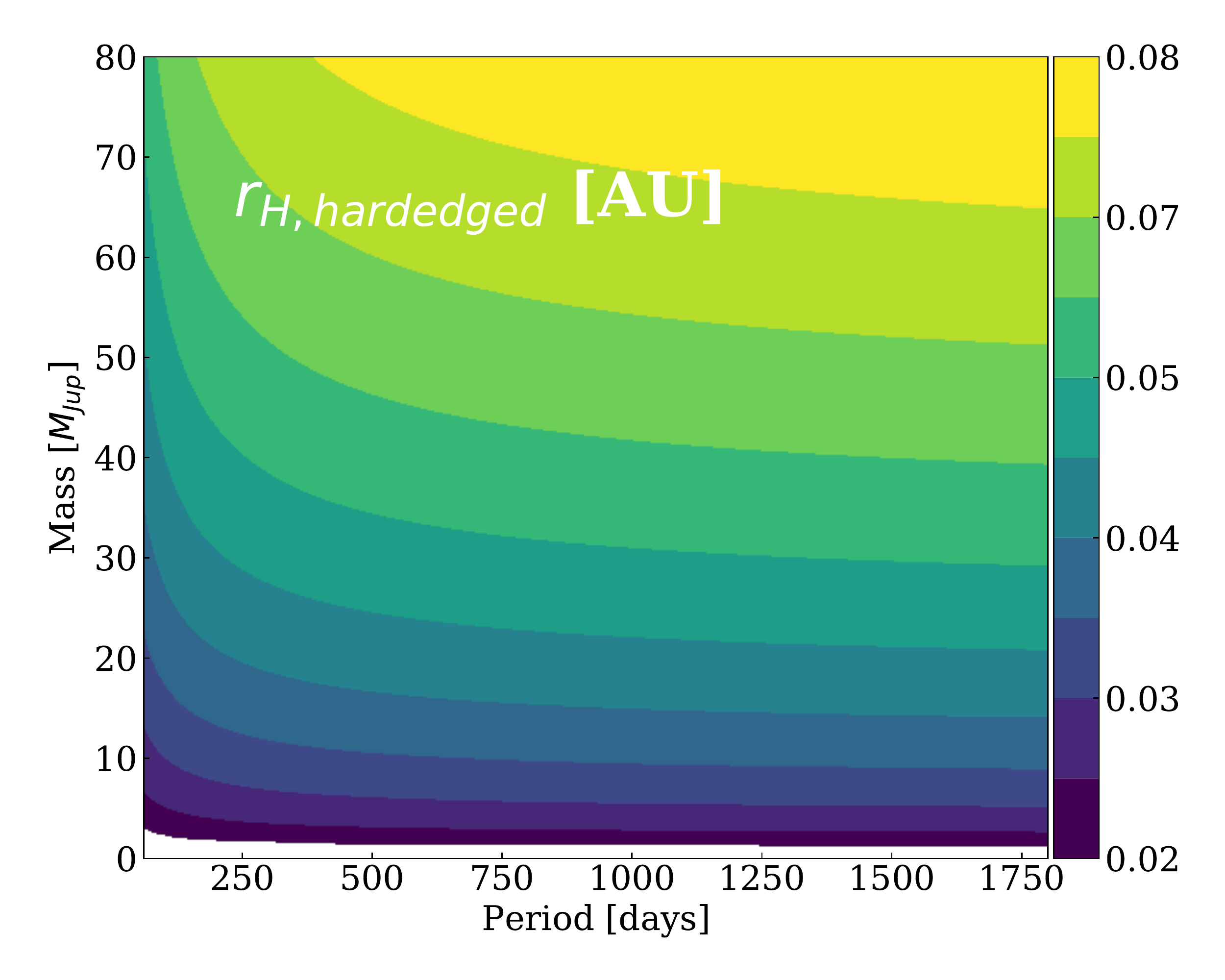} 
          \label{fig:rhhardedged1800}
        \end{subfigure}
        \begin{subfigure}{.4\textwidth}
          \centering
          \includegraphics[width=\linewidth]{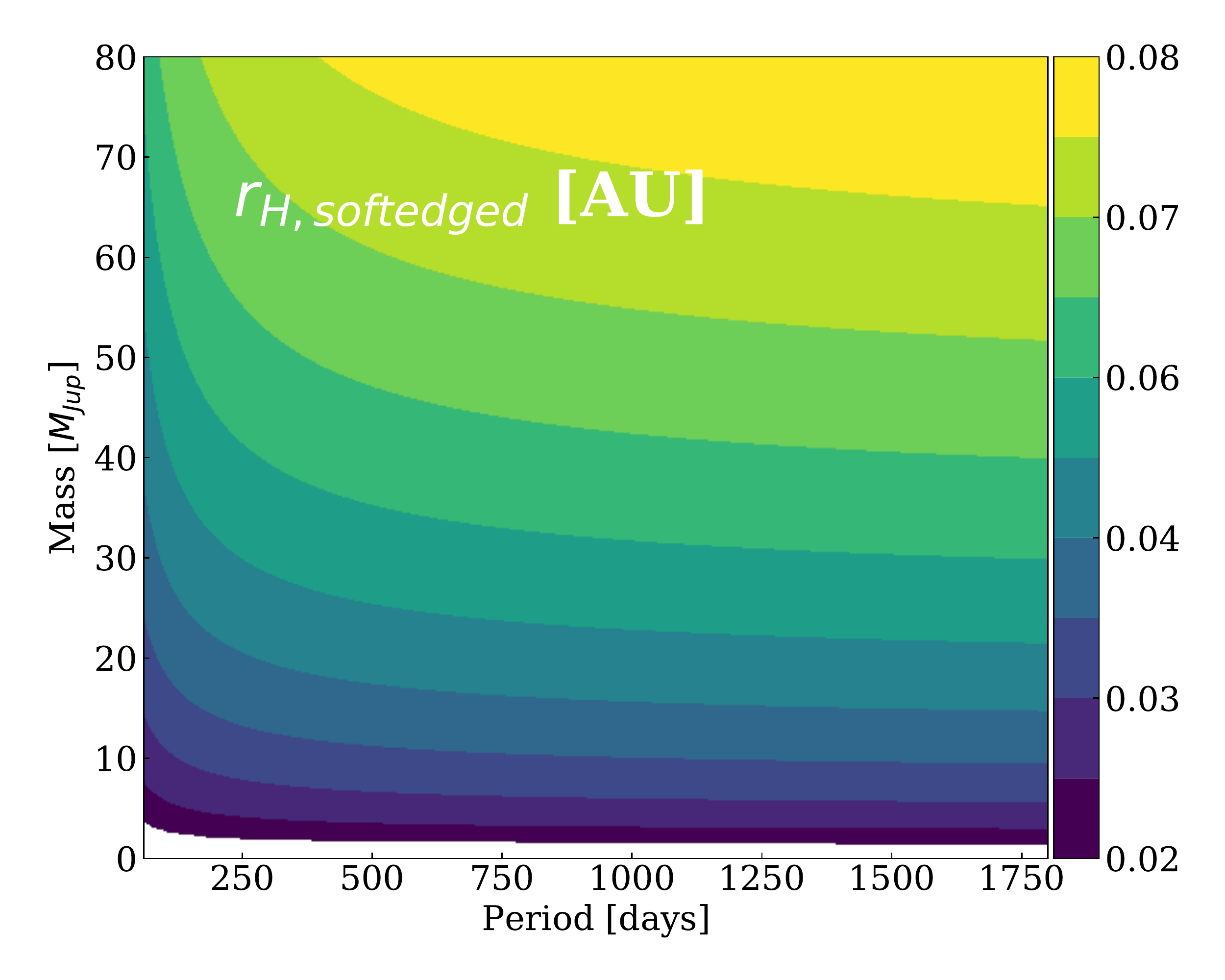}  
          \label{fig:rhsoftedged1800}
        \end{subfigure}
        \caption{ Parameter space maps for a period between 60 and 1800 days and mass between 0 and 80 $M_{\mathrm{Jup}}$.
          The left side is the hard edged disk, the right side is the soft edged disk.
          The white space at the bottom of the plots is from the constraint $R_d < 0.3$ $r_H$.}
        \label{fig:orbitalanalysis1800}
    \end{center}
\end{figure*}

The minimum mass is about 1.5 $M_{\mathrm{Jup}}$, set by the Hill sphere stability criterion ($R_{disk} < 0.3 \, R_{Hill}$) during the periastron passage of the disk.
For the most likely period of 290.230 days, the eccentricity of the orbit is around $e =$ 0.72, but as can be seen in Figure \ref{fig:orbitalanalysis1800}, the minimum eccentricity for the system can be as low as $e =$ 0.36. 
Longer orbital periods require higher eccentricities in order to keep the velocity at periastron equal to the derived transverse velocity.

Using the most likely periastron distance we found, and the effective temperature of the star, we can make an estimate of the equilibrium temperature of the disk.
For a periastron distance of 0.23 AU, corresponding to a period of about 290.230 days, the equilibrium of a fully absorbing disk around a companion orbiting \ac{etwo} would be about 420 K.
Using the second period of 235 days would result in a lower periastron distance and thus a higher equilibrium temperature, of about 430 K.
These temperatures exclude the possibility of a disk made of ice, so we conclude the disk is made of silicates.

\section{Discussion}
\label{sec:discuss}
\subsection{The EPIC 2202 Eclipser}

We have found two likely models of disks that could be responsible for the eclipse of \ac{etwo}, a fully opaque disk and an opaque disk with a soft edge.
Taking the simpler of the two models, the hard edged model gives a simple template that can be improved upon by adding more degrees of freedom, such as adding scattering or modeling an exponentially thinning disk.
We have found a most likely period for the system based on time-series photometry from ground-based surveys, but it does not cover the whole time domain, which, combined with the short duration of the eclipse likely means that we are missing other eclipses that could determine the period of the system.

\subsection{A Population of Small Disk Occulters}

There have been two other cases of circumsecondary disk occulters similar in size to \ac{etwo}.
The first is a disk around EPIC 2043 \citep{rappaport_deep_2019}, and the second is a disk around the binary star system V928~Tau \citep{van_dam_asymmetric_2020}.
A comparison between the radii of these stars as well as the parameters of the disk model obtained can be seen in Table \ref{tab:comparison}, and a comparison of the three systems with disk models can be seen in Figure \ref{fig:modelcomparison}.
Note that for \ac{etwo}, Table \ref{tab:comparison} and Figure \ref{fig:modelcomparison} show the simpler hard edged model.

\bgroup

\begin{table}[htb!]
    \centering
    \caption{Best Fit Values for the Small Disk Occulter Systems.}
    \begin{center}
        \begin{tabular}{lccc}
            \toprule
            \toprule
             Parameter                   & \ac{etwo}$^a$ & EPIC 2043 & V928 Tau$^b$  \\ 
            \midrule
             $R_*$ {[}$R_{\odot}${]}     & 0.83          & 0.63      & 1.38          \\
             $R_d$ {[}$R_*${]}           & 1.37          & 4.20      & 0.99          \\
             $R_d$ {[}$R_{\odot}${]}     & 1.14          & 2.60      & 1.39          \\
             $b$ {[}$R_*${]}             & -0.74         & -0.88     & -0.25         \\
             $i$ {[}$^\circ${]}          & 77.0          & 77.8      & 56.8          \\
             $\phi$ {[}$^\circ${]}       & -36.8         & 18.1      & 41.2          \\
             $v_t$ {[}km s$^{-1}${]}     & 77.4          & 38.0      & 73.5          \\
             Min. $e$ {[}-{]}            & 0.36          & 0.33      & 0.30          \\
             Min. $r_{peri}$ {[}AU{]}    & 0.17          & 0.13      & 0.10          \\
            \bottomrule
        \end{tabular}
        \label{tab:comparison}
    \end{center}
    \begin{tablenotes}
        \item{a.} Values from hard edged model. Note that $b$ and $\phi$ are flipped for easier comparison between the systems.
        \item{b.} Values taken around the primary star.
      \end{tablenotes}
\end{table}
\egroup

All of these systems have an inclination, tilt, and non-zero impact parameter, which produces the asymmetric eclipse.
Compared to the EPIC 2043 and V928 Tau systems, the companion of \ac{etwo} has the smallest disk in absolute size.
Nevertheless, all of these disks are larger than the size expected for Roche rings, where the tidal disruption radius is considerably smaller for the assumed masses.
In terms of disk size relative to the star, the disk around \ac{etwo} lies between EPIC 2043 and V928 Tau.

The detection of these systems is biased by the method used to find them - asymmetric eclipses with depths of tens of percent, in stars with flat or slowly varying light curves due to astrophysical activity.
If the disks are close to edge on, then their photometric signal may well be very small and therefore undetectable. 
If they have significant inclination but no tilt, then they present time symmetric light curves which might be misidentified as transiting stellar objects or grazing transits.
Having a less than 90$^\circ$ inclination with a tilt between 10$^\circ$ and 70$^\circ$, and a non-zero impact parameter to break symmetry, results in an asymmetric light curve that can be visually identified.
We propose that similar systems are present in the \textit{Kepler} and \textit{K2} light curves, but they have not been identified as such.
Identification of these disks through photometry from the ground is especially challenging, given the typically short duration of these eclipses, on the timescales of hours or less.

All three companions are required to be in an eccentric orbit around the star, with a minimum eccentricity of about 0.3, in order to explain the derived transverse velocity and the lack of other eclipses within the \textit{K2} data set.
There are on the order of 200 planets detected with eccentricities greater than 0.3 and with orbital periods between 100 and 1000 days\footnote{e.g. see \url{https://exoplanetarchive.ipac.caltech.edu/cgi-bin/TblView/nph-tblView?app=ExoTbls&config=PS}} so these orbital parameters are not unusual, although higher measured eccentricities are more rare for multiple exoplanet systems \citep{2015PNAS..112...20L}.
This minimum eccentricity could be explained by a perturbing companion that causes the companion with the inclined disk to be in an eccentric orbit - the V928 Tau system is a binary star system, but we are not currently aware of any companions around EPIC 2043 or EPIC 2202.
A peturbing companion may provide an explanation for the disks in the form of a planetessimal collision caused by a strong scattering event that put the companion into the observed eccentricity that we deduce.

All three light curves show a similar pattern of residuals - the points of ingress and egress show deviations from our model leading to the large $\chi^2$ values, and there is a difference in the deepest part of the eclipse, possibly as a result of the model not constraining enough degrees of freedom for an accurate model.
Especially in the case of \ac{etwo} and V928 Tau there is a distinct and similar down-up-down-up-down pattern in the residuals (see Figures \ref{fig:hardedgemodel}-\ref{fig:comparisonplot}).
For EPIC 2043 the residual pattern starts with down-up, but afterwards the pattern is less distinct.
The down-up-down-up-down pattern comes from the actual eclipse having less steep `wings' than the model allows for and a deeper minimum.
Both for \ac{etwo} and V928 Tau adding a soft edge to the model did not significantly improve the fit.
This alludes to a physical phenomenon which is not included in the model.
Future studies will include forward scattering and more complex structures for the putative disks.

\begin{figure*}[ht]
    \begin{center}
        \includegraphics[width=\linewidth]{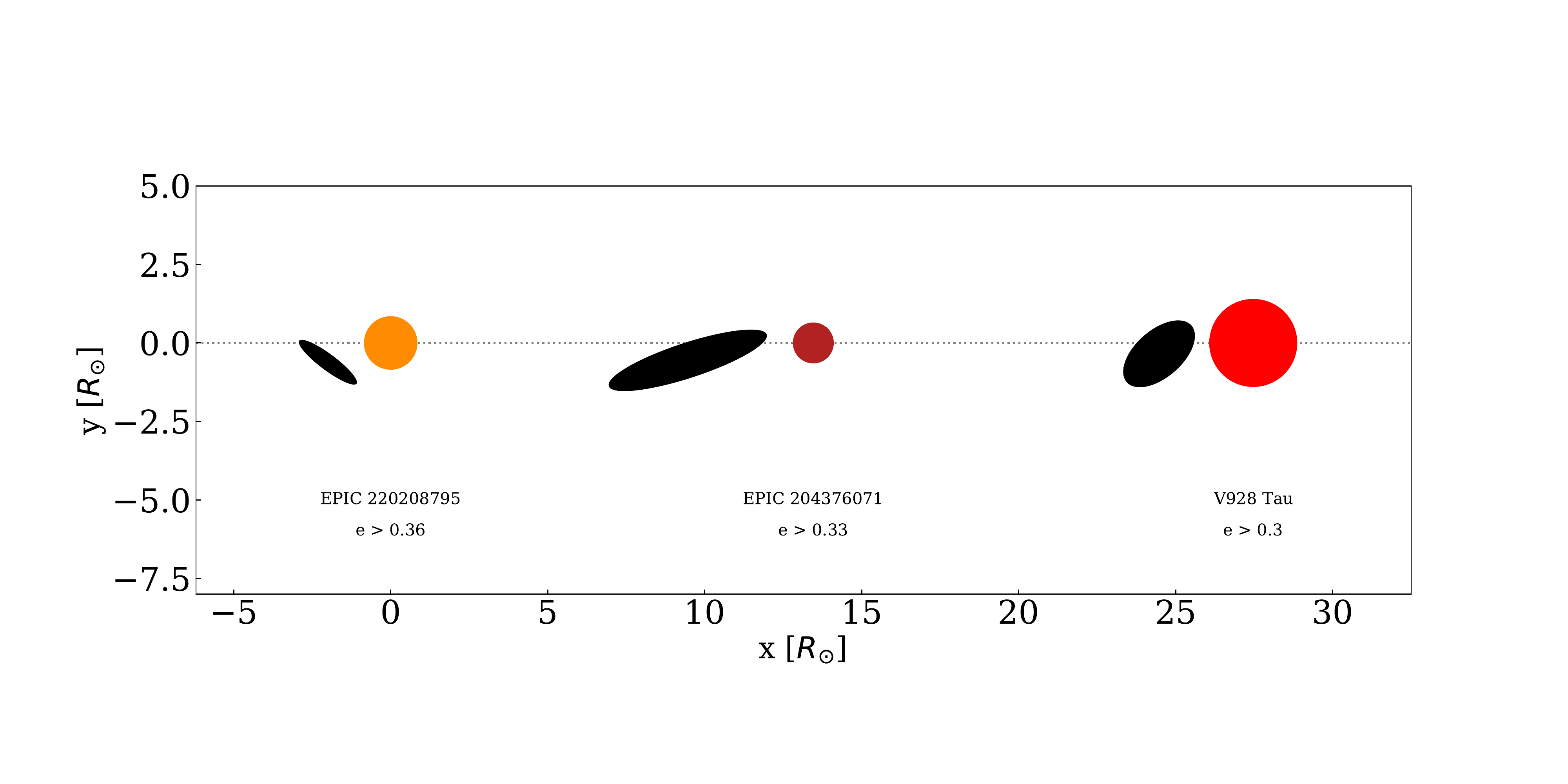}
        \vspace{-20pt}
        \caption{A comparison of three disk occulters.
          The disks move from left to right.
          The colours of the stars are based on their effective temperature relative to each other, where dark red is the coldest and orange is the hottest star.
          \ac{etwo} has been mirrored along a line through $y =$ 0 compared to the model obtained earlier as this has no effect on the shape of the light curve.}
        \label{fig:modelcomparison}
        \vspace{5pt}
    \end{center}
\end{figure*}

\section{Conclusions}\label{sec:conc}

We looked at the light curves initially found in \textit{K2} data by \citet{lacourse_single_2018}, from which we selected \ac{etwo} as the most likely candidate of a circumsecondary disk transit.
Using a modified version of \texttt{pyPplusS} in combination with \texttt{emcee}, we modeled the eclipse with a hard edged and soft edged disk.
Both of the models have similar residuals that the fitting did not account for.
This distinct down-up-down-up-down pattern in the residuals is also seen in the proposed inclined disk system of V928 Tau, and also partly in EPIC 2043.
The hard edged model is the simplest of the two, and so we keep our model of an opaque disk around a companion of \ac{etwo}, with a disk size of 1.37 $R_*$, an impact parameter of 0.74 $R_*$, an inclination of 77.01$^\circ$, a tilt of 36.81$^\circ$ and a transverse velocity of 77.4 km s$^{-1}$.
A period analysis using ground-based data of EPIC 2202 resulted in two most likely periods, namely, 235.587 and 290.230 days.
Taking the fitted transverse velocity we derived orbital parameters of the companion, such as a minimum eccentricity of $\sim$ 0.36.
EPIC 2043 and V928 Tau have roughly the same minimal eccentricity.
This eccentricity could be caused by a perturbing companion, but for EPIC 2043 and \ac{etwo} it is not known if there is such a companion. 

Future research includes constraining the period of the companion by detecting other eclipses, or by spectroscopic monitoring to determine the mass, or upper limit on the mass, of the secondary companion.
The star is an early K dwarf star, which is more amenable to spectroscopic radial velocity measurements, as the other two systems are around M dwarf stars.
If we assume that the periastron velocity was observed during the eclipse, we can estimate the amplitude of the radial velocity curve to be on the order of 70 km s$^{-1}$  $(M_P/M_*)\approx 250$ m s$^{-1}$, which would be possible to observe with a spectrograph on an 8m-class telescope.
The eclipse duration is only 7.2 hours, which makes it difficult to detect from ground-based observatories subject to diurnal window functions, but using the predictions for next eclipses based on the found periods, there could be observations planned for these dates to constrain the period of the companion.
An observational campaign for the American Association of Variable Star Observers (AAVSO) would be a likely path for detecting subsequent eclipses, although the magnitude of \ac{etwo} might leave it as observable for larger aperture telescopes.
Direct imaging and high contrast observations of the star could see if there is a perturbing companion around EPIC 2043 or \ac{etwo}, as well as observations in filters other than \textit{K2}'s $K_p$ filter.
Observations during future eclipses will help determine the specific composition and grain size distribution of the disk and companion.
Taking spectra of the star and then comparing its absorption features to the absorption features of the disk could tell us about any chemical reactions taking place in the disk.
The model can be extended to include scattering, an exponentially thinning disk or more rings with variable opacity to improve upon the model and find out what the origin is of the pattern in the residuals.

Determining the orbital period of these three systems, and finding more such systems within current and future wide field photometric surveys will give us an idea of their occurrence statistics.
Ideally, finding the orbital period of one of these systems will enable a spectroscopic campaign that can characterise the makeup and chemistry within these disks.

\begin{acknowledgements}
  
This research has used the SIMBAD database, operated at CDS, Strasbourg, France \citep{wenger2000}.
This work has used data from the European Space Agency (ESA) mission {\it Gaia} (\url{https://www.cosmos.esa.int/gaia}), processed by the {\it Gaia} Data Processing and Analysis Consortium (DPAC, \url{https://www.cosmos.esa.int/web/gaia/dpac/consortium}).
Funding for the DPAC has been provided by national institutions, in particular the institutions participating in the {\it Gaia} Multilateral Agreement.
To achieve the scientific results presented in this article we made use of the \emph{Python} programming language\footnote{Python Software Foundation, \url{https://www.python.org/}}, especially the \emph{SciPy} \citep{virtanen2020}, \emph{NumPy} \citep{numpy}, \emph{Matplotlib} \citep{Matplotlib}, \emph{emcee} \citep{foreman-mackey2013}, and \emph{astropy} \citep{astropy_1,astropy_2} packages.
This paper includes data collected by the \textit{Kepler} mission and obtained from the MAST data archive at the Space Telescope Science Institute (STScI).
Funding for the \textit{Kepler} mission is provided by the NASA Science Mission Directorate. STScI is operated by the Association of Universities for Research in Astronomy, Inc., under NASA contract NAS 5–26555.
We thank the Las Cumbres Observatory and its staff for its continuing support of the ASAS-SN project, and the Ohio State University College of Arts and Sciences Technology Services for helping us set up and maintain the ASAS-SN variable stars and photometry databases.
ASAS-SN is supported by the Gordon and Betty Moore Foundation through grant GBMF5490 to the Ohio State University and NSF grant AST-1515927.
Development of ASAS-SN has been supported by NSF grant AST-0908816, the Mt. Cuba Astronomical Foundation, the Center for Cosmology and AstroParticle Physics at the Ohio State University, the Chinese Academy of Sciences South America Center for Astronomy (CASSACA), the Villum Foundation, and George Skestos.
Part of this research was carried out in part at the Jet Propulsion Laboratory, California Institute of Technology, under a contract with the National Aeronautics and Space Administration (80NM0018D0004).
This publication makes use of VOSA, developed under the Spanish Virtual Observatory project supported by the Spanish MINECO through grant AyA2017-84089.
VOSA has been partially updated by using funding from the European Union's Horizon 2020 Research and Innovation Programme, under Grant Agreement number 776403 (EXOPLANETS-A) 

\end{acknowledgements}

\bibliography{references,BRPref}

\end{document}